\documentclass{aa}  

\usepackage{graphicx}
\usepackage{txfonts}
\usepackage{lipsum}
\usepackage{subcaption}
\usepackage{lscape}
\usepackage{placeins}

\begin{document}

\title{The shortest detected intra-day variability of active galactic nuclei in TESS survey}

\author{Heechan Yuk\inst{1}
        \and
        Xinyu Dai\inst{1}
        \and
        Natalie Kovacevic\inst{1}
        }

\institute{Homer L. Dodge Department of Physics and Astronomy,
          University of Oklahoma, 
          440 W. Brooks St., Norman, OK 73019, USA
          }

\date{March 00, 2026}

  \abstract
   {Active galactic nuclei (AGNs) are known to be variable in almost all wavelengths and timescales. The shortest variability timescale of AGNs can be used to probe the smallest scale structures within AGNs.}
   {We aim to measure the shortest detected variability timescale, $t_{\textrm{min,ul}}$, of type 1 radio-quiet Seyfert galaxies and analyse their characteristics.}
   {We extracted Transiting Exoplanet Survey Satellite (TESS) light curves of 47 Seyfert 1 galaxies. We measured the power spectral densities (PSDs) of the sample, modelled by a power law model plus a constant noise, and constrained the shortest detected AGN variability timescale as the power law component exceeds the constant noise and systematic uncertainties indicated by the upper limits of non-variable quiescent galaxies' PSDs.}
   {We measured the upper limits of the shortest variability timescale to be $\log(t_{\textrm{min,ul}}/\textrm{hrs})=0.85\pm0.55$.  We compared these upper limits to a range of theoretical AGN variability timescales, and the natural interpretation of our measured $t_{\textrm{min,ul}}$ is the light crossing scale from a coherently varying region, where the measured $t_{\textrm{min,ul}}$ corresponds to the range from a few to thousands of gravitational radii. A significant fraction of these light crossing scales is smaller than the accretion disk emission sizes measured by quasar microlensing, reverberation mapping, or theoretical accretion disk models. Since we only measure the upper limits, the true physical shortest variability timescales are even shorter. We measure the power law index to be $\alpha=2.0\pm0.2$, and find weak anticorrelations with the black hole mass and luminosity.}
   {Our analysis suggests that the shortest optical variability is driven by a compact region smaller than the accretion disk size, potentially by X-ray reprocessing. Alternatively, this shortest timescale variability suggests that the accretion disk can be inhomogeneous potentially caused by turbulence from magnetorotational instability or magnetic reconnections.}


\maketitle
\nolinenumbers

\section{Introduction}

Active galactic nuclei (AGNs) are observed to be variable throughout the entire electromagnetic spectrum in almost all timescales. AGN variability provides useful insights about the inner structures of AGNs. One example is the light crossing scale. As light travels at a finite speed, the variability timescale can be scaled by the speed of light to measure the size of the coherently varying region emitting at that wavelength. For example, AGN variability was first observed in timescale of days, suggesting that the size of an AGN is in order of the Solar system. This led to the idea of AGN being powered by accretion of material onto a compact object, namely a supermassive black hole (SMBH; \citealt{lyndenbell69}). The light crossing scale has been routinely applied to different wavelengths to constrain AGN emission sizes to the first order in optical, IR, X-ray bands \citep[e.g.][]{fabian15,pozonunez15,jha22}. Another example is reverberation mapping; by measuring the time lag between brightness fluctuations in different wavelengths, which corresponds to different regions in the AGN structure, observers can estimate the size of the broad line region (BLR) of the AGN and the mass of the SMBH \citep[e.g.][]{peterson93}. Studying the power spectral densities (PSDs) of AGNs also provides an alternative method to probe the SMBH. AGN PSDs are often modelled with a broken power law, and the break frequency, the location where the power law index shifts, is observed to be correlated with the black hole mass \citep[e.g.][]{kelly09,macleod10,gonzalezmartin18,burke21,tarrant25}. 

Variability observed in timescales shorter than a day, or intra-day, intra-night, or microvariability, is important in AGN studies, as it can be used to probe the smallest structures of AGNs. The SMBH or its innermost stable orbit might provide the shortest timescale, but if variability arises from accretion disk inhomogeneities, the observed variability timescale can be even smaller. Potentially, there can be no lower limit on the variability timescale, and thus the measurement of shortest AGN variability time scale can provide an important insight on the source of AGN variability.

The short time scale AGN variability in X-rays is well studied because of the larger AGN variability amplitudes in this band and the photon counting nature of X-ray observations. The X-ray AGN PSD studies report that variability is detected in frequencies upwards of $10^{-4}$ Hz to $10^{-2}$ Hz, or in timescale of minutes to hours \citep[e.g.][]{uttley02,markowitz03,mchardy04,gonzalezmartin12,gonzalezmartin18,rani25}. This would suggest that the X-ray emitting region is in order of a few $r_g$, which is consistent with independent measurements from microlensing \citep[e.g.][]{morgan08,dai10}.

Optical AGN variability studies are often limited by the fact that the data mostly comes from ground-based telescopes, combined with the smaller intra-day variability amplitudes, and, as a result, high-cadence monitoring data of AGN is rare and measurements of intra-day variability can be challenging. Many previous intra-day optical AGN variability involved using statistical methods, such as F-test, to detect variability and measuring the variability amplitude and duty cycle, the fraction of time when the AGN is varying \citep[e.g.][]{romero99,goyal12,goyal13,kumar15}. Different classes of AGNs display different duty cycles--greater duty cycles were reported for AGNs with higher radio loudness and optical polarization--suggesting different physical conditions \citep[e.g.][]{romero99,goyal13}. Also, many of these studies were conducted on blazars or radio-loud AGNs, where the variability is strongly affected by the relativistic jet, which makes them less ideal for understanding the innermost structure of AGNs.

The advent of high-cadence long-term space-based timing surveys opened a new window for intra-day optical AGN variability studies. The Kepler Space Telescope, which made observations with cadence in order of minutes over many years, enabled optical PSD studies similar to X-ray studies \citep[e.g.][]{aranzana18,smith18}, where variability in timescales of hours to days was detected. Transiting Exoplanet Survey Satellite (TESS; \citealt{ricker15}) is another space-based high-cadence survey mission. While Kepler is limited to a specific region in the celestial sphere, covering approximately 116 square degrees, TESS is an all-sky survey, providing the ideal data set for the variability study of a general AGN population. Recent intra-day AGN variability studies using the TESS data involve blazars and radio-loud AGNs \citep[e.g.][]{dingler24,yang24}. In order to study the short timescale optical variability near the central engine, we focus on examining TESS light curves of radio quiet AGNs.

In this study, we seek to measure the upper limits of the shortest AGN variability timescale, $t_{\textrm{min,ul}}$, that is detectable in the TESS survey. As mentioned earlier, observations reveal that the AGN PSDs are modelled by a single or broken power law \citep[e.g.][]{mchardy04,gonzalezmartin18}. By measuring the frequency where the power law component dominates over noise or systematics, we can measure the shortest detected variability timescale. We introduce our sample and data in Section \ref{secdata} and describe methods in Section \ref{secmethods}. We present our results in Section \ref{secresults}, discuss the results' implications in Section \ref{secdisccussion}, and conclude in section \ref{secconclusion}. Throughout this paper, we assume the $\Lambda$CDM cosmological model with $\Omega_{\Lambda}=0.7$, $\Omega_M=0.3$, and $H_0=70$ km s$^{-1}$ Mpc$^{-1}$.

\section{Sample and Data}
\label{secdata}

We examine a sample of bright Seyfert galaxies with at least one TESS sector observed. Our sample is primarily derived from the AGN sample of \citet{gonzalezmartin18} and the Swift BAT 9-month AGN catalogue \citep{tueller08}. We limit our sample to type 1, as type 2 Seyfert galaxies are observed to display very weak to non-detectable TESS variability \citep[e.g.][]{kovacevic25}. We follow the classification of the respective catalogue. In order to exclude radio-loud AGNs, we used the radio loudness parameter $\mathcal{R}=F_{\textrm{5GHz}}/F_B$, where $F_{\textrm{5GHz}}$ and $F_B$ are flux density in 5 GHz and $B$-band, respectively. AGNs with $\mathcal{R}>10$ are classified as radio-loud AGNs \citep[e.g.][]{chiaberge11}. We retrieved photometry information from NASA/IPAC Extragalactic Database (NED)\footnote{https://doi.org/10.26132/NED1}. The final sample size is 47 and they are listed on Table \ref{tesssecs}.

For this sample, we also gathered the black hole mass from literature. The black hole masses reported by \citet{koss17} using the single-epoch H$\beta$ emission do not include uncertainties. Instead, the authors mention that this method is known to have systematic uncertainties of about 0.3 dex, which we use as uncertainties. The black hole masses were not reported for a few objects. For these objects, we retrieved V-band magnitudes of these targets from All-Sky Automated Survey for SuperNovae (ASAS-SN; \citealt{shappee14,kochanek17}), estimated the $I$-band spheroid luminosity, and used the spheroid luminosity scaling relation of \citet{bennert21}, as described in \citet{yuk22}.

TESS is a satellite designed to scan a portion of the sky with high cadence, for approximately 27 days at a time. After each 27-day cycle, called a sector, it points at a different position, eventually covering the entire celestial sphere. It is equipped with four cameras, each with 24\textdegree\ by 24\textdegree\ field of view, observing 24\textdegree\ by 96\textdegree\ simultaneously. TESS filter spans from 600 to 1000 nm. The primary mission, which began with launch in 2018 until sector 26 in July 2020, took full frame images (FFIs) every 30 minutes. Following the primary mission was the first extended mission, which lasted until sector 55 in September 2022 with a 10-minute cadence FFIs. Since then, TESS entered the second extended mission, which is observing with a 200-second cadence. The typical limiting magnitudes are approximately 19.6 for 30-minute FFIs, 19.0 for 10-minute FFIs, and 18.4 for 200-second FFIs. For this study, we limit our data to primary and first extended missions.

We extracted the TESS light curves using the image subtraction method. First, the FFIs are cut into 750 pixel by 750 pixels postage stamps around the target. Then the first 100 good images are combined into a reference image. This reference image is subtracted from all postage stamps. Then we performed a point spread function photometry on the subtracted images to measure the flux relative to the reference image. More detailed information on the TESS light curve extraction method can be found in  \citet{fausnaugh21} and \citet{vallely21}.

Once a light curve is extracted, we examined along with light curves of nearby pixels for any anomalies, potentially caused by the instrument or a nearby source affecting the galaxy's light curve. Some examples of anomalies can be found in \citet{yuk25}.

We also extracted TESS light curves of quiescent galaxies in order to measure the systematic variability. These galaxies are selected from the Two-Micron All Sky Survey (2MASS; \citealt{skrutskie06}) extended sources catalogue \citep{jarrett00}, that are bright ($K<11.34$) and away from the galactic plane ($|b|>15$ \textdegree). In addition, we required the galaxy sample to fall in the magnitude range of the Swift-BAT 9-month AGN sample and have comparable average flux uncertainties. In total, we analysed TESS light curves of 17 galaxies to examine the systematics. The galaxies' flux distribution is $\log(F/\textrm{mJy})=0.96\pm0.38$, while the AGNs' flux distribution is $\log(F/\textrm{mJy})=1.06\pm0.43$.

\section{Methods}
\label{secmethods}

TESS light curves are sampled mostly in regular intervals. However, in the middle of a sector is a discontinuity for approximately a day for data transmission. Also, removing bad data points or outliers can introduce additional gaps. So, we use the Lomb-Scargle periodogram \citep[e.g.][]{lomb76,scargle82} to construct the PSDs from TESS light curves, which is ideal for computing the PSDs for unevenly sampled light curves. We computed the PSD within the frequency range of $1/T$ to $1/(2\Delta T_{\textrm{samp}})$, where $T$ is the duration of a sector ($\sim$27 days) and $\Delta T_{\textrm{samp}}$ is the sampling interval (30 minutes for sectors $\le$26 and 10 minutes for sectors $>$26). We then binned the PSDs by a factor of 1.6, and took the standard deviation of the mean for each bin as the uncertainty. The Lomb-Scargle periodogram method was previously used to successfully measure AGN PSDs using TESS data \citep[e.g.][]{burke21,yuk25}.

Although AGN PSDs are commonly measured as a broken power law, the typical break frequency is in order of $10^{-3}$ to $10^{-2}$ day$^{-1}$ \citep[e.g.][]{burke21,tarrant25,yuk25}, which is below the range of TESS frequencies. Some studies report high-frequency breaks in the TESS PSD regime \citep[e.g.][]{burke20,yuk25}, and these high-frequency breaks are at the low frequency part of the TESS PSDs, which affect little to the measurement of PSDs at high frequencies. We chose to use a simple power law model to fit the TESS PSD to simplify the analysis:

\begin{equation}
P(\nu) = A\bigg(\frac{\nu}{\nu_{\textrm{ref}}}\bigg)^{-\alpha}+C,
\label{powerlaw}
\end{equation}
where $P$ is the power, $A$ is the amplitude at $\nu_{\textrm{ref}}$, $\nu_{\textrm{ref}}$ is the reference frequency (set to the median value of TESS $\nu$ range), and $C$ is the white noise level. Once the model is fit, theoretically, the shortest variability timescale can be calculated by computing the frequency where the power law component exceeds the noise, $\nu_{\textrm{max}} = \nu_{\textrm{ref}}(C/A)^{-1/\alpha}$. We tested the broken power law model in the discussion and yielded consistent $\nu_{\textrm{max}}$ values.

\begin{figure*}
    \centering
    \includegraphics[width=0.8\linewidth]{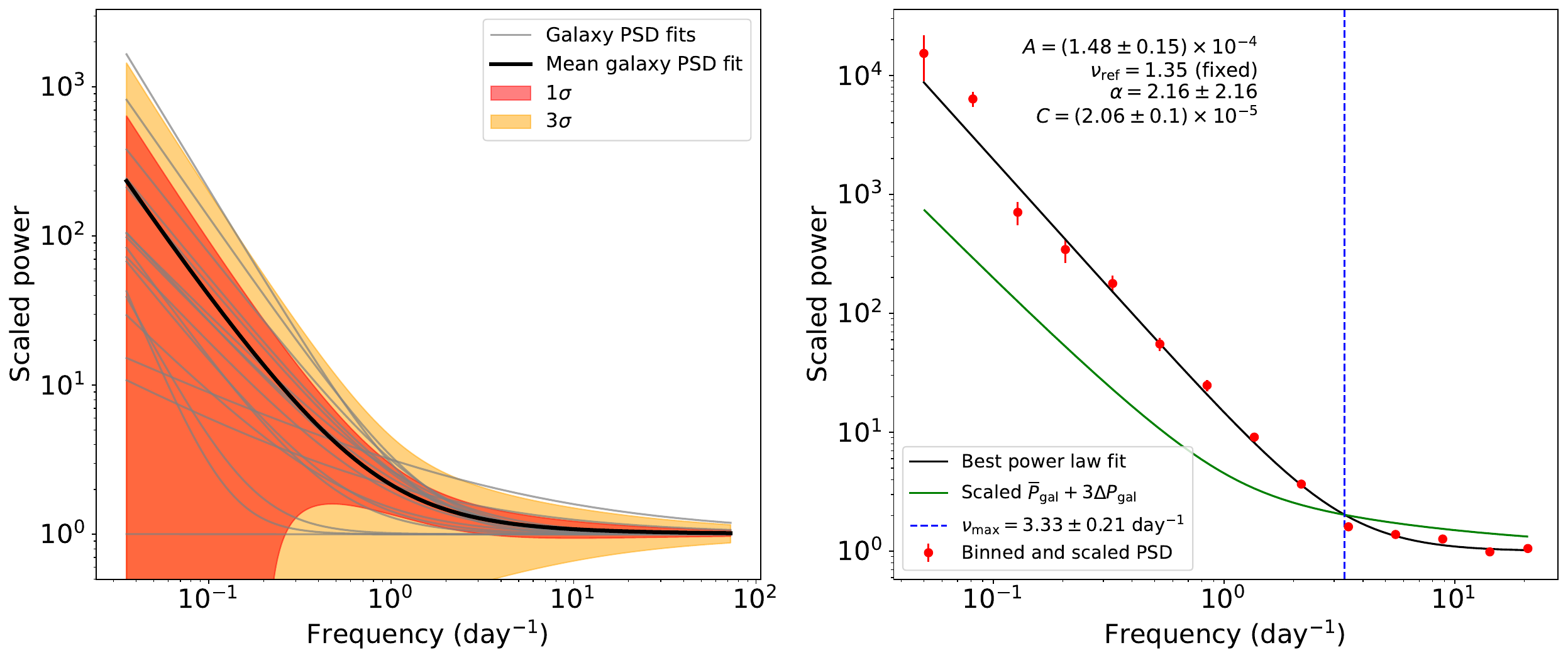}
    \caption{Left: The systematics of TESS survey and light curve extraction technique indicated by the PSDs of non-variable galaxy light curves, where the y-axis is the scaled power relative to the constant white noise. Each gray line indicates the power law plus white noise fit for an individual galaxy's light curve, the black line represents the mean galaxy PSD, and the red and yellow shaded areas represent the 1-sigma and 3-sigma limits, respectively.
    Right: Comparison between Seyfert 1 AGN, ESO 362-G018, Sector 5 PSD, which has the measured $t_{\textrm{min,ul}}$ value equal to the median of the AGN sample, and the upper limits of galaxies as a representative to illustrate the measurement of the shortest detected variability time scale. The text next to the PSD lists the parameters for the best fit. PSDs in both panels are normalized by their respective noise levels.}
    \label{comparetogal}
\end{figure*}

However, it is possible for systematic variability to be present in the TESS light curves. To examine it, we reduced TESS light curves of several quiescent galaxies, computed PSDs, and fit the power law plus white noise model. We found that $\alpha\sim1.5$ for systematic variability measured in quiescent galaxies and $\alpha\sim2.0$ for AGNs. This suggests that TESS light curves of galaxies can show a red noise spectrum due to systematic effects; however, the slope is shallower than the red noise spectrum of AGN and more importantly, the amplitudes for galaxy spectra are smaller, which enable us to detect AGN variability beyond these noises. In particular, we measured the mean and the standard deviation of the PSD with respect to the white noise level for the quiescent galaxies' PSDs. We estimated the upper limit of the systematic variability as the mean plus 3-sigma of the galaxies' PSDs. We then located where the AGN PSDs exceed this upper limit to measure the shortest variability timescale. An example of comparing galaxies' average PSD and an AGN PSD is shown on Figure \ref{comparetogal}. For AGNs with multiple TESS sectors, we computed $\nu_{\textrm{max}}$ for each sector separately. We found a scatter in $\nu_{\textrm{max}}$ measurements among different sectors, suggesting that there exist sector-to-sector and pixel-to-pixel dependent instrumental factors. Due to our sample size constraints, we are unable to characterise these dependences. So, for this sub-sample of AGNs, we took the largest $\nu_{\textrm{max}}$ as our shortest detected variability timescale $t_{\textrm{min,ul}}$. As the measured $t_{\textrm{min,ul}}$ values from this method is in observed frame, we corrected it by a factor of $(1+z)$ in order to calculate the rest-frame timescale for further analysis.

\begin{figure*}
    \centering
    \includegraphics[width=0.8\linewidth]{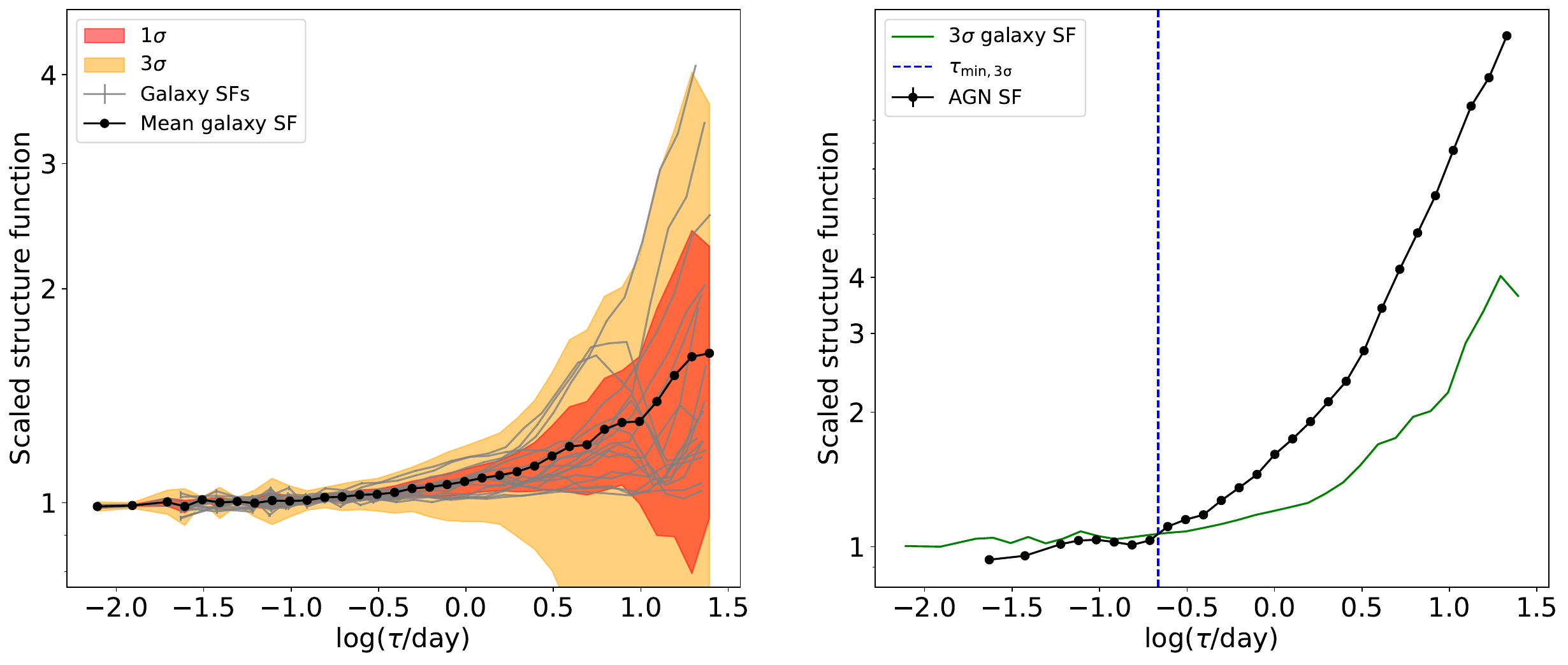}
    \caption{Left: The distribution of quiescent galaxies' structure functions, normalized by the noise level, similar to Figure \ref{comparetogal} left. Right: Comparison of structure functions of galaxies and NGC 931 sector 18 to demonstrate the smallest variability timescale measurement.}
    \label{comparetogal_sf}
\end{figure*}

In order to complement the timescale measurements, we also computed the structure function. To calculate the structure function, we used the following definition:
\begin{equation}
    \textrm{SF}(\tau) = \sqrt{\langle[m(t-\tau)-m(t)]^2\rangle-\sigma_{\textrm{noise}}^2},
\end{equation}
where $\tau$ is the time lag between two observations, $m$ is the magnitude, and 
\begin{equation}
    \sigma_{\textrm{noise}}^2 = \langle\sigma_{\textrm{err}}^2(t)+\sigma_{\textrm{err}}^2(t+\tau)\rangle
\end{equation}
is the uncertainty correction where $\sigma_{\textrm{err}}$ is the measurement uncertainty in magnitude \citep[e.g.][]{bauer09,middei17}. The AGN structure functions are measured with an increasing trend at long timescales as expected; however, we found that the power law with white noise model fits poorly to structure functions. So, for each structure function, we estimated the noise level to be the average of the smallest 25\% of the structure function, which typically locates at the shortest timescale segment, and normalized the structure function by this noise estimate. The normalized structure functions of galaxies and AGNs were then compared to measure $t_{\textrm{min,ul}}$ (Figure \ref{comparetogal_sf}).

\section{Results}
\label{secresults}

\begin{figure}
    \centering
    \includegraphics[width=\linewidth]{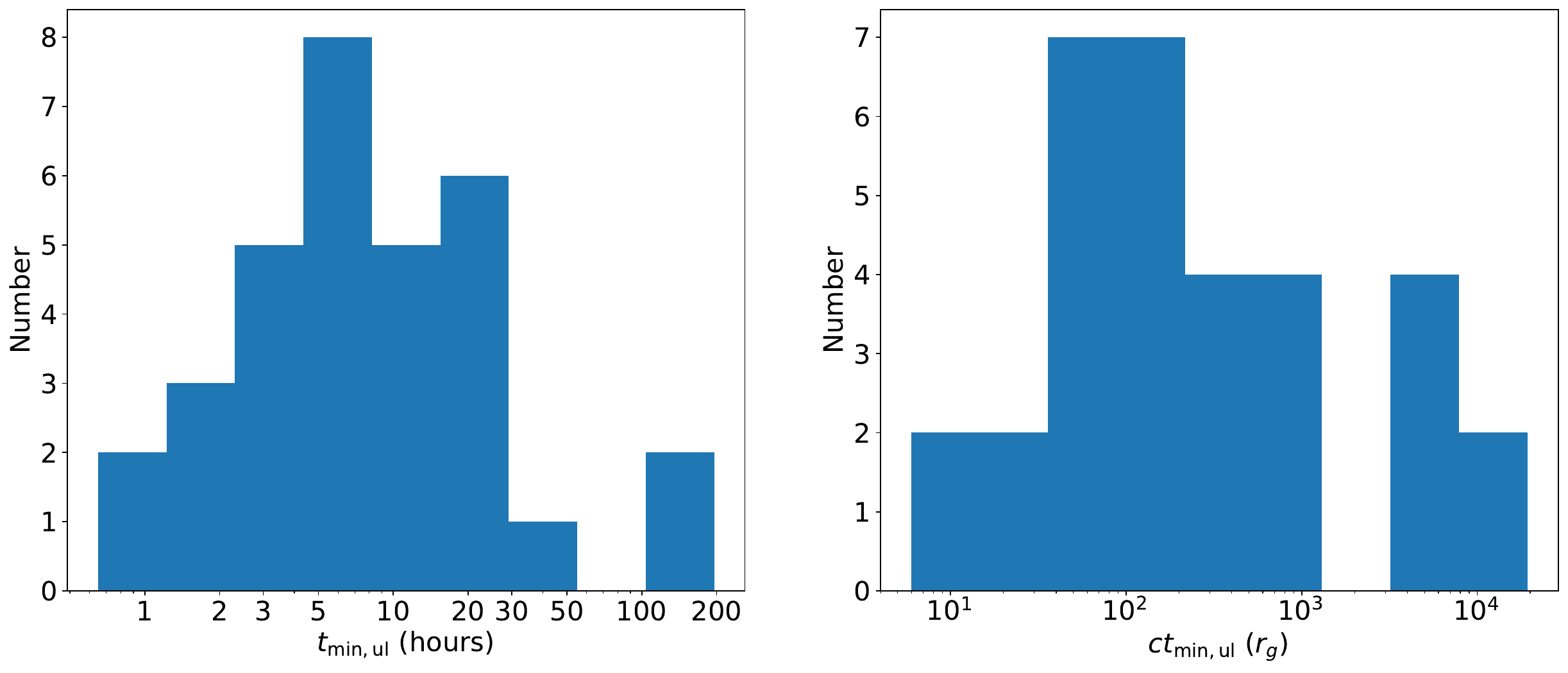}
    \caption{Left: Distribution of the shortest detected TESS variability timescales of the sample. Right: Histogram showing the distribution of $c t_{\textrm{min,ul}}$ in terms of $r_g$.}
    \label{timescaleshist}
\end{figure}

Out of the 47 AGNs, we measured the shortest detected variability timescales for 32. We measured that the $\nu_{\textrm{max}}$ ranges from $1.2\times10^{-1}$ to $3.5\times10^1$ day$^{-1}$, with the median at $3.4$ day$^{-1}$. Conversely, the shortest variability timescale detected by TESS light curves ranges from 0.68 to 193 hours, with the median at 7.1 hours, or $\log(t_{\textrm{min,ul}}/\textrm{hrs})=0.85\pm0.55$. The distribution of $t_{\textrm{min,ul}}=1/\nu_{\textrm{max}}$ is shown on Figure \ref{timescaleshist}. We found that the remaining 15 AGNs have small variability amplitudes with the power component lower than the systematics.

\begin{figure}
    \centering
    \includegraphics[width=\linewidth]{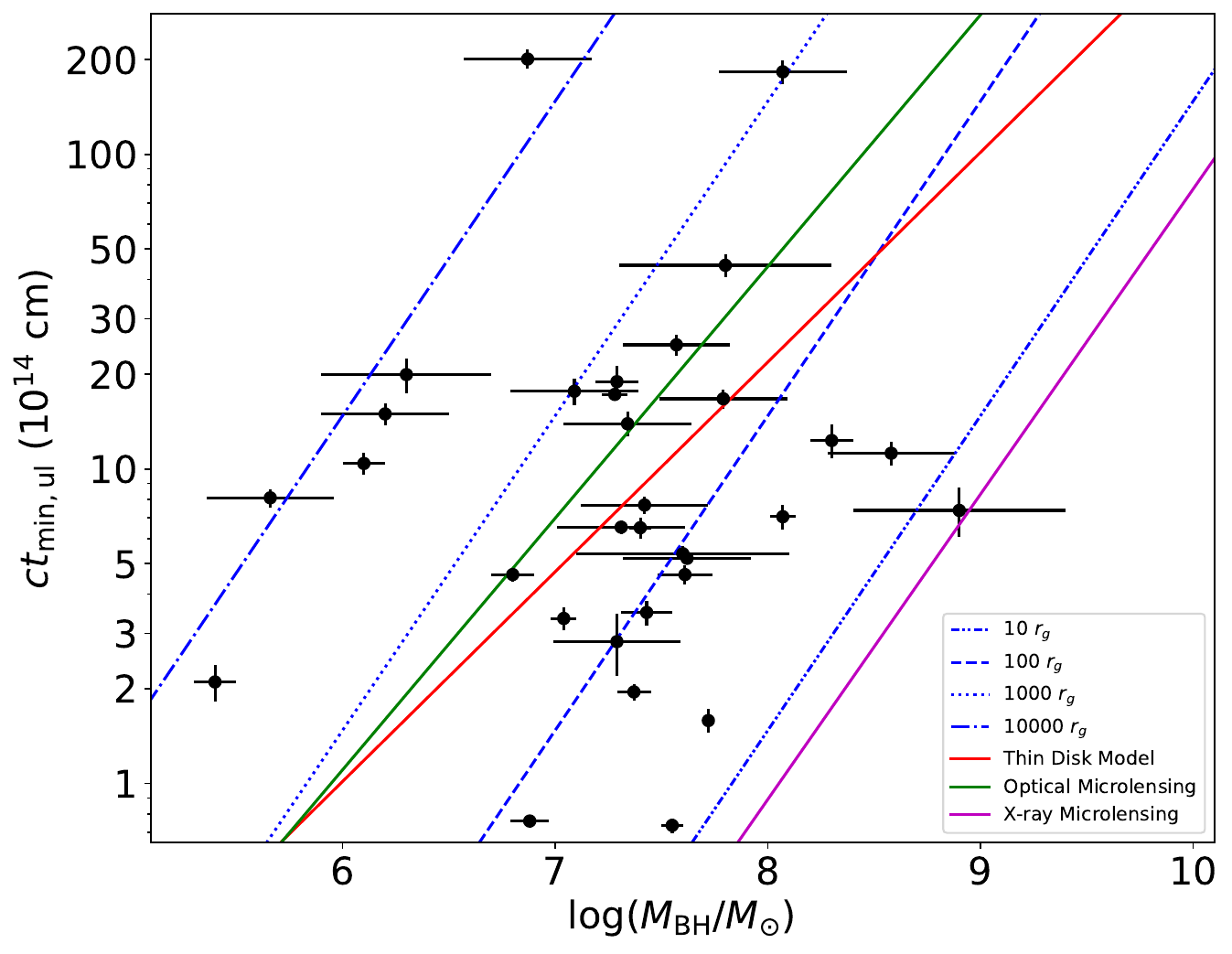}
    \caption{Distribution of the shortest detected TESS variability timescales of the sample, plotted against the black hole mass. The blue dashed lines indicate the different values of gravitational radii for the given black hole mass, and the green and red solid lines indicate the theoretical and measured accretion disk and corona sizes from literature \citep{shakura73,morgan10,dogruel20}.}
    \label{timescales}
\end{figure}

The light crossing scale is the most natural explanation for these short time scale variations. We compare the light crossing scale corresponding to the shortest variability timescale to the gravitational radius of the black holes, $r_g = GM/c^2$, quasar microlensing accretion disk size measurements, and the theoretical and measured accretion disk scaling relations with black hole mass in Figure \ref{timescales}. These light crossing scales correspond to a few to thousands $r_g$, and for massive black holes, we can probe the variability down to a few $r_g$ scale (Figure \ref{timescales}). The quasar microlensing accretion disk size of \citet{morgan10} is 

\begin{equation}
    R_{2500}=10^{15.78\pm0.12}\bigg(\frac{M_{\textrm{BH}}}{10^9M_{\odot}}\bigg)^{0.80\pm0.17}\textrm{ cm}.
\end{equation}
While \citet{morgan10} measured disk size at rest-frame UV, we scale the size to the TESS band by $R \propto \lambda^{4/3}$. The X-ray emitting region size for radio-quiet quasars by \citet{dogruel20} is

\begin{equation}
    R_{\textrm{X,RQ}} = 10^{6.19\pm3.06}\bigg(\frac{M_{\textrm{BH}}}{M_{\odot}}\bigg)^{0.97\pm0.35}\textrm{ cm.}
\end{equation}
The thin disk theoretical model prediction is:

\begin{equation}
    R_{\lambda_{\textrm{rest}}}=
    9.7\times10^{15}
    \bigg(\frac{\lambda_{\textrm{rest}}}{\mu \textrm{m}}\bigg)^{4/3}
    \bigg(\frac{M_{\textrm{BH}}}{10^9 M_{\odot}}\bigg)^{2/3}
    \bigg(\frac{L}{\eta L_E}\bigg)^{1/3} 
    \textrm{cm},
\end{equation}
where $\lambda_{\textrm{rest}}$ is the rest wavelength, $L/L_E$ is the Eddington ratio, and $\eta$ is the accretion efficiency \citep[e.g.][]{shakura73,morgan10}. As our sample is local bright Seyfert galaxies with low redshifts, we approximated $\lambda_{\textrm{rest}}$ as the central wavelength of TESS ($\sim$780 nm), the Eddington ratio as $L/L_E=0.3$, and the accretion efficiency as $\eta=0.1$. 

Out of 32 AGNs with $t_{\textrm{min,ul}}$ measurements, we find that 17 AGNs' $ct_{\textrm{min,ul}}$ are consistent or above the accretion disk size from the thin disk model by \citet{shakura73}, 15 are consistent or above the microlensing accretion disk size measurement of \citet{morgan10}, and all are larger but only one is consistent with the microlensing measurement of X-ray emitting region size by \citet{dogruel20}. 47\% to 53\% of the sample display variability in scales smaller than the size of the accretion disk in optical wavelength from thin disk prediction or microlensing measurements, respectively. This suggests that the source of the optical variability is smaller in size, some approaching the size of the X-ray corona.

\begin{figure*}
    \centering
    \includegraphics[width=0.8\linewidth]{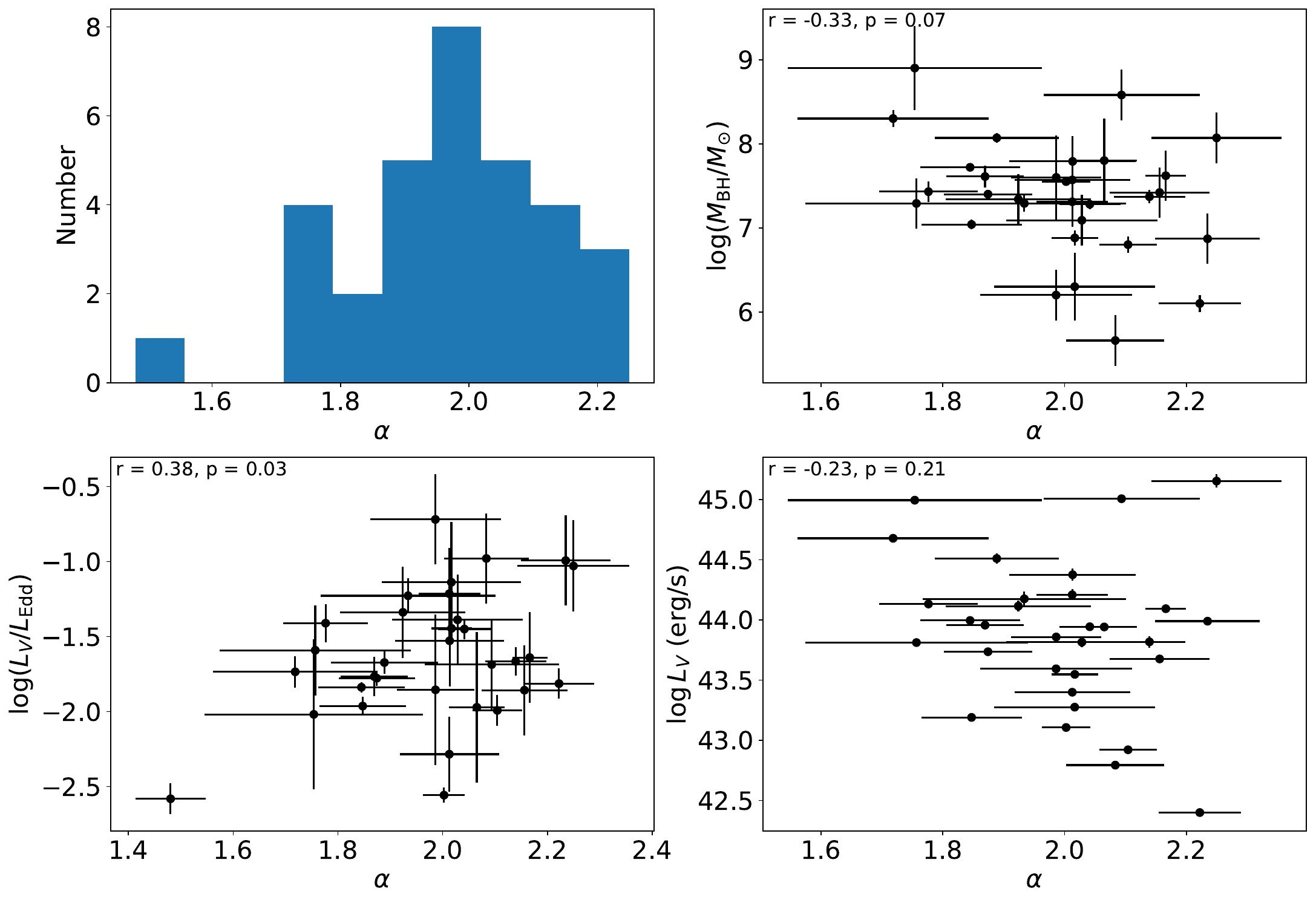}
    \caption{Top left: Distribution of the power law indices for the PSD power law fit. 
    Black hole mass (top right), $L_V/L_{\textrm{Edd}}$ (bottom left), and V-band luminosity (bottom right) plotted against the power law index.}
    \label{correlations}
\end{figure*}

\begin{figure*}
    \centering
    \includegraphics[width=0.8\linewidth]{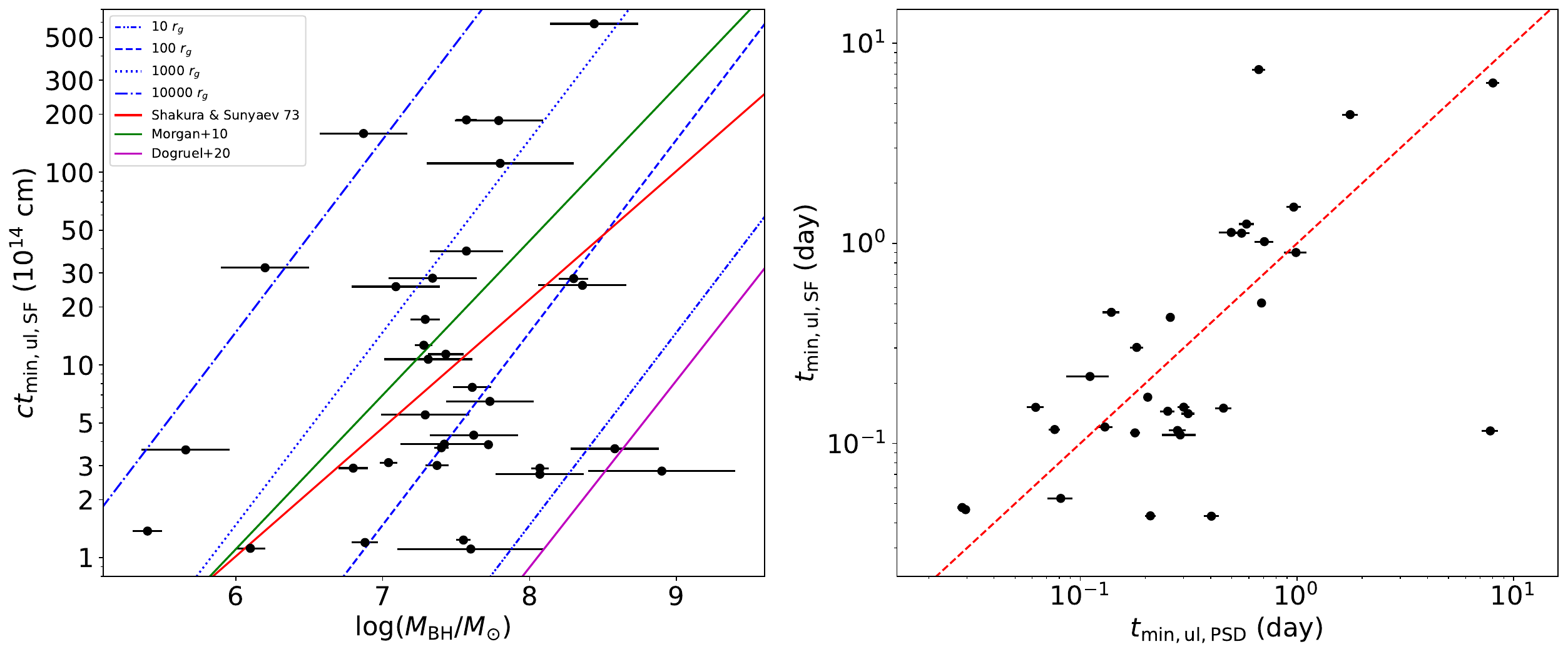}
    \caption{Left: Same as Figure \ref{timescales}, but with $t_{\textrm{min,ul}}$ measured from structure functions. Right: Comparison between $t_{\textrm{min,ul}}$ measured in the observed frame from PSDs and structure functions. The red dashed line represents the one-to-one relation.}
    \label{sfresults}
\end{figure*}

We measured the power law indices to be $\alpha=2.0\pm0.2$. This is consistent with the damped random walk model (DRW; \citealt[e.g.][]{kelly09,kozlowski16}) at the high frequency regime. We note that Kepler PSD studies, which probes similar timescales, report steeper high-frequency power law index measurements \citep[e.g.][]{mushotzky11,smith18}, which requires further investigation for confirmation. We discovered that power law index is weakly correlated with pseudo-Eddington ratio ($r=0.38$, $p=0.03$), weakly anticorrelated with the black hole mass ($r=-0.33$, $p=0.07$), and not correlated with luminosity ($r=-0.23$, $p=0.21$) (Figure \ref{correlations}).  The significances of these weakly correlated relations are low, and a larger sample is needed to further confirm these correlations.

By using structure functions, we measured $t_{\textrm{min,ul}}$ for 35 out of 47 AGNs, with the median and spread to be $\log(t_{\textrm{min,ul}}/\textrm{hrs})=0.72\pm0.71$, which is consistent with the measurements from PSDs. The fraction of $t_{\textrm{min,ul}}$ consistent or smaller than the predicted or measured accretion disk size range from 43\% to 51\%, and when the two sets of $t_{\textrm{min,ul}}$ measurements are compared, they approximately match with the one-to-one relation (Figure \ref{sfresults}).

\section{Discussion}
\label{secdisccussion}

As mentioned in Section \ref{secmethods}, we used a single power law, rather than a broken power law, to model the TESS PSDs, because previously measured break frequencies are outside the TESS frequency range or close to the lower limit. To check if our assumption that the effect of high-frequency breaks on the shortest variability timescale measurements is negligible, we tested by fitting the AGN PSDs with the broken power law model with the high-frequency breaks measured by \citet{yuk25}. We found that the shortest variability timescale measurements using single power law and broken power law are consistent with each other.

\begin{figure*}
    \centering
    \includegraphics[width=0.8\linewidth]{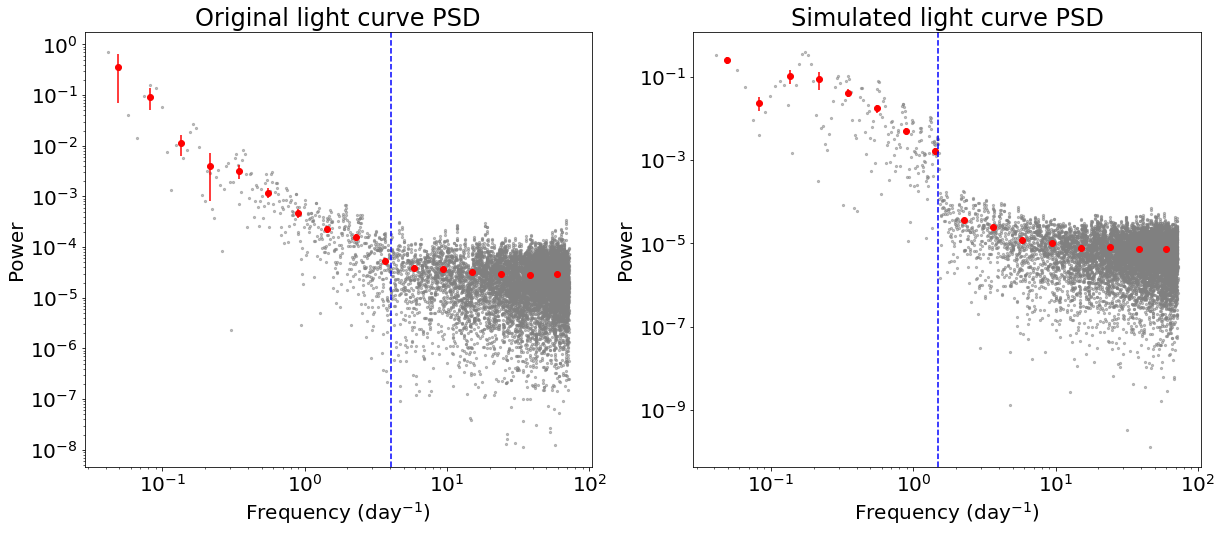}
    \caption{Left: The PSD of an actual TESS light curve (NGC 3783 sector 36). The blue dashed line indicates the shortest detected variability timescale. Right: The PSD of a simulated light curve with $\alpha=1.8$ and $\nu_{\textrm{c}}=1.5$ day$^{-1}$. The blue dashed line indicates $\nu_{\textrm{c}}$.}
    \label{simcompare}
\end{figure*}

We measured the shortest AGN optical variability timescale by modelling the TESS PSD as the power law plus a constant noise, where the power law component is detected above the noise and the systematics. Thus the truly shortest variability timescale might be even shorter under the noise level. For targets with multiple TESS sectors, we used the shortest $t_{\textrm{min,ul}}$ value in the subsequent analysis, which might be subject to sector-to-sector and pixel-to-pixel instrumental and observational systematics, and future studies with much larger AGN samples can potentially better address these issues. However, it is also possible that the true shortest variability timescale of the sample is greater than our measured values and our measurements are affected by PSD bleeding. So, we examine simulated light curves with an underlying PSD of cutoff power law, where the power is zero--or no variability--past the cutoff frequency, where the cutoff frequency is lower than our detected $\nu_{\textrm{max}}$. This simulates the scenario where no variability is present below the set timescale. Using the method of \citet{timmer95}, we simulated light curves with the following PSD:

\begin{equation}
P(\nu) \propto \begin{cases}
    \nu^{-\alpha} & \text{if } \nu<\nu_{\textrm{c}} \\
    0, & \text{if } \nu\ge\nu_{\textrm{c}} \\
\end{cases}
\end{equation}
where $P$ is the power, $\alpha$ is the power law index, and $\nu_{\textrm{c}}$ is the cutoff frequency. We simulated continuous light curves, applied observations windows of TESS, and added measurement noises based on typical errorbar sizes of TESS light curves. Figure \ref{simcompare} shows an example of PSDs of an actual TESS light curve and a simulated light curve. For the simulated PSD, a discontinuity at the cutoff frequency is clearly visible, which is not seen in the observed PSD measurements.

\begin{figure}
    \centering
    \includegraphics[width=\linewidth]{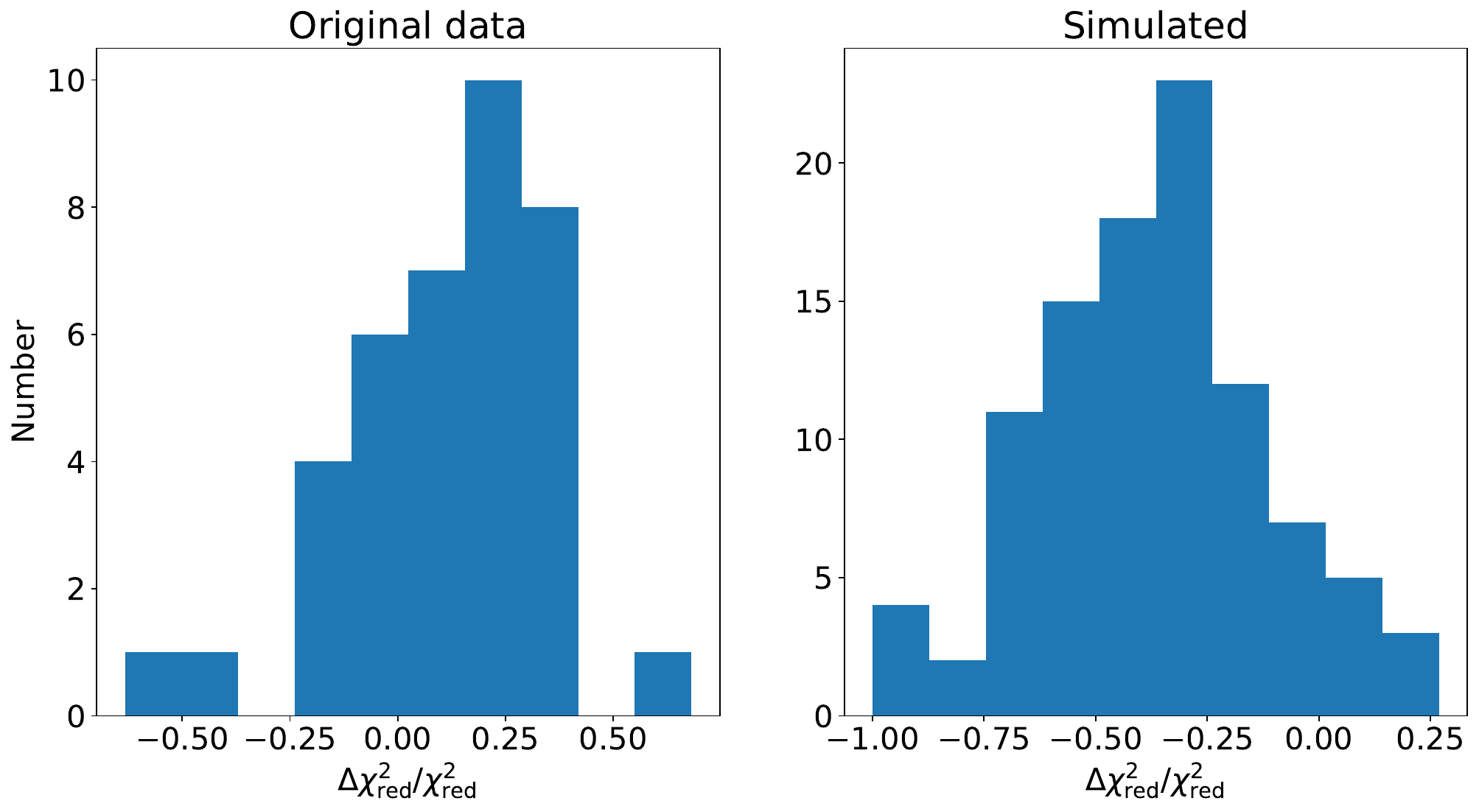}
    \caption{The distributions of $\Delta\chi^2/\chi^2$ between simple power law model and the discontinuous power law model fits for original data (left) and the simulated data (right).}
    \label{chi2dist}
\end{figure}

To quantify the difference between the two PSD shapes, we simulated 100 light curves with $\nu_c=1.5$ day$^{-1}$ and $\alpha=1.8$ and fit two models to real and simulated PSDs and computed the difference in $\chi^2$. The first model is a simple power law plus white noise model (Equation \ref{powerlaw}), and the second model is a discontinuous power law,

\begin{equation}
P(\nu) = \begin{cases}
    A_1(\frac{\nu}{\nu_{\textrm{c}}})^{-\alpha_1}+C & \text{if } \nu<\nu_{\textrm{c}} \\
    A_2(\frac{\nu}{\nu_{\textrm{c}}})^{-\alpha_2}+C, & \text{if } \nu\ge\nu_{\textrm{c}} \\
\end{cases}
\end{equation}
where $A_1$ and $A_2$ are amplitudes at $\nu_{\textrm{c}}$ with $A_1\ge A_2$, $\alpha_1$ and $\alpha_2$ are power law indices, $\nu_{\textrm{c}}$ is the cutoff frequency, and $C$ is the white noise term. This model is designed to detect a discontinuity in the PSD. If the PSD is continuous, $A_1$ should approach $A_2$ and $\alpha_1$ should approach $\alpha_2$, so the fitting results and $\chi^2$ should be similar to fitting to Equation \ref{powerlaw}. The distributions of $\Delta\chi^2/\chi^2$ is shown on Figure \ref{chi2dist}. For PSDs retrieved from the actual data, the difference peaks at zero, suggesting that those PSDs are continuous. However, for PSDs from simulated light curves, $\Delta\chi^2/\chi^2$ peaks at about $-0.4$, meaning $\chi^2$ improves by about 40\% by using the discontinuous power law, suggesting that there exist a significant discontinuity in simulated PSDs. With such clear differences between the two distributions, we can confirm that the shortest variability timescale we measured is real.

\begin{table*}[]
\caption{List of various timescales associated with AGN variability}
\centering
\begin{tabular}{lll}
\hline\hline
Category & Formula & Timescale for $10^7 M_{\odot}$ \\ \hline
Light-crossing$^1$ & $t_{\textrm{lc}}\simeq0.011(\frac{M_{\textrm{BH}}}{10^7M_{\odot}})(\frac{R}{10R_S})$ days & 0.18 day \\
Free-fall$^2$ & $t_{\textrm{ff}}\simeq0.046(\frac{M_{\textrm{BH}}}{10^7M_{\odot}})(\frac{R}{10R_S})^{3/2}$ days & 2.9 days \\
Orbital$^1$ & $t_{\textrm{orb}}\simeq0.33(\frac{M_{\textrm{BH}}}{10^7M_{\odot}})(\frac{R}{10R_S})^{3/2}$ days & 21 days \\
Hydrostatic equilibrium$^3$ & in the same order as orbital timescale &  \\
Sound-crossing in vertical direction$^3$ & in the same order as orbital timescale &  \\
Thermal$^1$ & $t_{\textrm{th}}\simeq0.53(\frac{\gamma}{0.1})^{-1}(\frac{M_{\textrm{BH}}}{10^7M_{\odot}})(\frac{R}{10R_S})^{3/2}$ days & 34 days \\
Sound-crossing in radial direction$^4$ & $t_{\textrm{sound,R}}\simeq0.53(\frac{R}{10H})(\frac{M_{\textrm{BH}}}{10^7M_{\odot}})(\frac{R}{10R_S})^{3/2}$ days & $1.7\times10^2$ days \\
Viscous$^4$ & $t_{\textrm{visc}}\simeq53(\frac{R}{10H})^2(\frac{\gamma}{0.1})^{-1}(\frac{M_{\textrm{BH}}}{10^7M_{\odot}})(\frac{R}{10R_S})^{3/2}$ days & $8.2\times10^4$ days \\
Cold disk removal$^3$ & $t_{\textrm{evap}}\simeq3.7\times10^2(\frac{L}{0.1L_E})(\frac{M_{\textrm{BH}}}{10^7M_{\odot}})(\frac{R}{10R_S})^2$ days & $2.8\times10^5$ days \\ \hline
\end{tabular}
\tablefoot{$^1$\citet{edelson99}; $^2$\citet{smith18}; $^3$\citet{czerny06}; $^4$\citet{paolillo25}. We use the thin disk model accretion disk size by \citet{shakura73} for $R$ and estimate the ratio of thickness to radius of the accretion disk as $\frac{H}{R}\simeq12.5(\frac{R}{R_S})^{-1}(\frac{L}{L_E})(1-\sqrt{\frac{3R_S}{R}})$ and the viscosity parameter as $\gamma=0.1$.}
\label{timescalestable}
\end{table*}

There is a number of timescales associated with optical variability. We computed the various timescales, taking the radial distance from the SMBH, $R$, as the theoretical thin disk model accretion disk size (Table \ref{timescalestable}). We found that the light-crossing timescale of the accretion disk from quasar microlensing measurement or thin disk prediction is the closest to the shortest variability timescale we measure from TESS light curves, in order of sub-day. The reverberation mapping disk sizes are largely consistent with quasar microlensing sizes \citep[e.g.][]{edelson15,jha22}. The free-fall timescale, which may represent the advection-dominated accretion flows (ADAF) timescale, is roughly in the order of magnitude as the longest variability timescales we measured. Other timescales, such as orbital and thermal timescales, are multiple orders of magnitude longer than our measurements.

Since light crossing scale measures a coherently varying emission regions, we find that a significant fraction of AGNs have the coherently varying region smaller than the accretion disk size from the thin disk model prediction or microlensing measurements. This suggests that the short timescale AGN variability may originate from regions smaller than the accretion disk. We introduce two potential explanations for this observation. Previous studies suggest that the sizes of X-ray and optical emitting regions in AGNs are in order of $10 r_g$ and $100 r_g$, respectively \citep[e.g.][]{morgan08,dai10}. For AGNs with the most massive black holes, the optical variability timescale corresponds to less than $10 r_g$. From this, we can hypothesize that the short timescale optical variability is driven by the inner X-ray emitting region near the accretion disk via reprocessing. The model of X-ray coronal emission being reflected from and reprocessed by the accretion disk has been explored in a number of previous studies \citep[e.g.][]{pounds90,george91,shappee14b,papoutsis24}. For AGNs with less massive black holes, it is possible that the variability timescale measurements are limited by the instrument's sensitivity more severely than more massive black holes. But AGNs with such small timescale length is limited to a small fraction of our sample. Most of our sample have the variability timescale length greater than the measured X-ray emitting region size. So, an alternative interpretation is that the optical variability originates from the inhomogeneities in the accretion disk. Some studies theorize that local inhomogeneities cause accretion disk instabilities that lead to AGN variability \citep[e.g.][]{kawaguchi98,trevese02,dexter12}. The physical origin can be turbulence induced by magnetorotational instability \citep[e.g.][]{balbus91,beckwith11} or magnetic reconnection due to various physics \citep[e.g.][]{ball18,ripperda20}.

Although we do not find a statistically significant correlation between the power law index and the luminosity, the visual inspection and the correlation coefficient hint at a potentially weak anticorrelation. Previous Kepler PSD studies report such anticorrelation \citep{smith18}. Other studies report that AGN variability amplitude is anticorrelated with black hole mass \citep[e.g.][]{lu01,papadakis04,macleod10} and luminosity \citep[e.g.][]{wang22,kovacevic25}, which can explain the anticorrelation we observed with the power law index assuming similar variability amplitudes.

We also note that a few narrow line Seyfert 1 galaxies are included in our sample. But, we do not find a significant distinction from other Seyfert 1 galaxies. We excluded the jet contribution mostly through the radio loudness parameter. However, there can be jetted AGN with radio loudness parameter below the threshold. We identified a few of them and their short variability can be from the jet. Most of the targets in our sample are truly non-jetted.

\section{Conclusion}
\label{secconclusion}

We examined TESS light curves of 47 type 1 radio-quiet Seyfert galaxies to examine their short timescale variability. We computed the PSDs of the sample and by comparing to the white noise and systematic variability measured by the upper limits of quiescent galaxies' PSDs, we measured the shortest detected variability timescale to be $\log(t_{\textrm{min,ul}}/\textrm{hrs})=0.85\pm0.55$, or, in units of $r_g$, $\log(ct_{\textrm{min,ul}}/r_g)=2.31\pm0.88$. We scaled our variability timescales by the speed of light and compared them to theoretical and measured accretion disk size, and discovered that our light crossing scales are typically smaller.

AGNs exhibiting optical variability in such short timescales can be interpreted in two main ways. As such short timescale correlates to small lengths, comparable or smaller than the theoretical and observed accretion disk size, the optical variability may be driven by small scale sources. One possibility is that the X-ray corona drives the optical variability by reprocessing. Alternatively, inhomogeneities in the accretion disk due to magnetic turbulence or reconnection could cause the detected short timescale optical variability. This study demonstrates that short-term variability analysis of AGNs can provide a unique way to study the spatial properties of AGNs and the origin of the variability.

We measure the power law indices to be $\alpha=2.0\pm0.2$ and detect weak anticorrelations with the black hole mass and luminosity, which is consistent with previous studies.

\begin{acknowledgements}
We thank the anonymous referee for helpful feedback throughout the review process. We would like to acknowledge NASA funds 80NSSC22K0488, 80NSSC23K0379 and NSF fund AAG2307802.
\end{acknowledgements}

\bibliographystyle{aa}
\bibliography{reference}

@ARTICLE{yuk22,
       author = {{Yuk}, Heechan and {Dai}, Xinyu and {Jayasinghe}, T. and {Fu}, Hai and {Mishra}, Hora D. and {Kochanek}, Christopher S. and {Shappee}, Benjamin J. and {Stanek}, K.~Z.},
        title = "{Variability Selected Active Galactic Nuclei from ASAS-SN Survey: Constraining the Low Luminosity AGN Population}",
      journal = {\apj},
         year = 2022,
        month = may,
       volume = {930},
       number = {2},
          eid = {110},
        pages = {110},
          doi = {10.3847/1538-4357/ac6423},
archivePrefix = {arXiv},
       eprint = {2203.08348},
 primaryClass = {astro-ph.GA},
       adsurl = {https://ui.adsabs.harvard.edu/abs/2022ApJ...930..110Y}
}

@ARTICLE{gonzalezmartin18,
       author = {{Gonz{\'a}lez-Mart{\'\i}n}, Omaira},
        title = "{Update on the X-Ray Variability Plane for Active Galactic Nuclei: The Role of the Obscuration}",
      journal = {\apj},
         year = 2018,
        month = may,
       volume = {858},
       number = {1},
          eid = {2},
        pages = {2},
          doi = {10.3847/1538-4357/aab7ec},
archivePrefix = {arXiv},
       eprint = {1803.05925},
 primaryClass = {astro-ph.GA},
       adsurl = {https://ui.adsabs.harvard.edu/abs/2018ApJ...858....2G}
}

@ARTICLE{yuk25,
       author = {{Yuk}, Heechan and {Dai}, Xinyu},
        title = "{High-frequency breaks in the optical active galactic nucleus power spectral density}",
      journal = {\aap},
         year = 2025,
        month = jun,
       volume = {698},
          eid = {A105},
        pages = {A105},
          doi = {10.1051/0004-6361/202452469},
archivePrefix = {arXiv},
       eprint = {2306.17334},
 primaryClass = {astro-ph.HE},
       adsurl = {https://ui.adsabs.harvard.edu/abs/2025A&A...698A.105Y}
}

@ARTICLE{ricker15,
       author = {{Ricker}, George R. and {Winn}, Joshua N. and {Vanderspek}, Roland and {Latham}, David W. and {Bakos}, G{\'a}sp{\'a}r {\'A}. and {Bean}, Jacob L. and {Berta-Thompson}, Zachory K. and {Brown}, Timothy M. and {Buchhave}, Lars and {Butler}, Nathaniel R. and {Butler}, R. Paul and {Chaplin}, William J. and {Charbonneau}, David and {Christensen-Dalsgaard}, J{\o}rgen and {Clampin}, Mark and {Deming}, Drake and {Doty}, John and {De Lee}, Nathan and {Dressing}, Courtney and {Dunham}, Edward W. and {Endl}, Michael and {Fressin}, Francois and {Ge}, Jian and {Henning}, Thomas and {Holman}, Matthew J. and {Howard}, Andrew W. and {Ida}, Shigeru and {Jenkins}, Jon M. and {Jernigan}, Garrett and {Johnson}, John Asher and {Kaltenegger}, Lisa and {Kawai}, Nobuyuki and {Kjeldsen}, Hans and {Laughlin}, Gregory and {Levine}, Alan M. and {Lin}, Douglas and {Lissauer}, Jack J. and {MacQueen}, Phillip and {Marcy}, Geoffrey and {McCullough}, Peter R. and {Morton}, Timothy D. and {Narita}, Norio and {Paegert}, Martin and {Palle}, Enric and {Pepe}, Francesco and {Pepper}, Joshua and {Quirrenbach}, Andreas and {Rinehart}, Stephen A. and {Sasselov}, Dimitar and {Sato}, Bun'ei and {Seager}, Sara and {Sozzetti}, Alessandro and {Stassun}, Keivan G. and {Sullivan}, Peter and {Szentgyorgyi}, Andrew and {Torres}, Guillermo and {Udry}, Stephane and {Villasenor}, Joel},
        title = "{Transiting Exoplanet Survey Satellite (TESS)}",
      journal = {\jatis},
         year = 2015,
        month = jan,
       volume = {1},
          eid = {014003},
        pages = {014003},
          doi = {10.1117/1.JATIS.1.1.014003},
       adsurl = {https://ui.adsabs.harvard.edu/abs/2015JATIS...1a4003R}
}

@ARTICLE{timmer95,
       author = {{Timmer}, J. and {Koenig}, M.},
        title = "{On generating power law noise.}",
      journal = {\aap},
         year = 1995,
        month = aug,
       volume = {300},
        pages = {707},
       adsurl = {https://ui.adsabs.harvard.edu/abs/1995A&A...300..707T}
}

@ARTICLE{lomb76,
       author = {{Lomb}, N.~R.},
        title = "{Least-Squares Frequency Analysis of Unequally Spaced Data}",
      journal = {\apss},
         year = 1976,
        month = feb,
       volume = {39},
       number = {2},
        pages = {447-462},
          doi = {10.1007/BF00648343},
       adsurl = {https://ui.adsabs.harvard.edu/abs/1976Ap&SS..39..447L}
}

@ARTICLE{scargle82,
       author = {{Scargle}, J.~D.},
        title = "{Studies in astronomical time series analysis. II. Statistical aspects of spectral analysis of unevenly spaced data.}",
      journal = {\apj},
         year = 1982,
        month = dec,
       volume = {263},
        pages = {835-853},
          doi = {10.1086/160554},
       adsurl = {https://ui.adsabs.harvard.edu/abs/1982ApJ...263..835S}
}

@ARTICLE{burke21,
       author = {{Burke}, Colin J. and {Shen}, Yue and {Blaes}, Omer and {Gammie}, Charles F. and {Horne}, Keith and {Jiang}, Yan-Fei and {Liu}, Xin and {McHardy}, Ian M. and {Morgan}, Christopher W. and {Scaringi}, Simone and {Yang}, Qian},
        title = "{A characteristic optical variability time scale in astrophysical accretion disks}",
      journal = {Science},
         year = 2021,
        month = aug,
       volume = {373},
       number = {6556},
        pages = {789-792},
          doi = {10.1126/science.abg9933},
archivePrefix = {arXiv},
       eprint = {2108.05389},
 primaryClass = {astro-ph.GA},
       adsurl = {https://ui.adsabs.harvard.edu/abs/2021Sci...373..789B}
}

@ARTICLE{mchardy04,
       author = {{McHardy}, I.~M. and {Papadakis}, I.~E. and {Uttley}, P. and {Page}, M.~J. and {Mason}, K.~O.},
        title = "{Combined long and short time-scale X-ray variability of NGC 4051 with RXTE and XMM-Newton}",
      journal = {\mnras},
         year = 2004,
        month = mar,
       volume = {348},
       number = {3},
        pages = {783-801},
          doi = {10.1111/j.1365-2966.2004.07376.x},
archivePrefix = {arXiv},
       eprint = {astro-ph/0311220},
 primaryClass = {astro-ph},
       adsurl = {https://ui.adsabs.harvard.edu/abs/2004MNRAS.348..783M}
}

@ARTICLE{vallely21,
       author = {{Vallely}, P.~J. and {Kochanek}, C.~S. and {Stanek}, K.~Z. and {Fausnaugh}, M. and {Shappee}, B.~J.},
        title = "{High-cadence, early-time observations of core-collapse supernovae from the TESS prime mission}",
      journal = {\mnras},
         year = 2021,
        month = feb,
       volume = {500},
       number = {4},
        pages = {5639-5656},
          doi = {10.1093/mnras/staa3675},
archivePrefix = {arXiv},
       eprint = {2010.06596},
 primaryClass = {astro-ph.HE},
       adsurl = {https://ui.adsabs.harvard.edu/abs/2021MNRAS.500.5639V}
}

@ARTICLE{fausnaugh21,
       author = {{Fausnaugh}, M.~M. and {Vallely}, P.~J. and {Kochanek}, C.~S. and {Shappee}, B.~J. and {Stanek}, K.~Z. and {Tucker}, M.~A. and {Ricker}, George R. and {Vanderspek}, Roland and {Latham}, David W. and {Seager}, S. and {Winn}, Joshua N. and {Jenkins}, Jon M. and {Berta-Thompson}, Zachory K. and {Daylan}, Tansu and {Doty}, John P. and {F{\H{u}}r{\'e}sz}, G{\'a}bor and {Levine}, Alan M. and {Morris}, Robert and {P{\'a}l}, Andr{\'a}s and {Sha}, Lizhou and {Ting}, Eric B. and {Wohler}, Bill},
        title = "{Early-time Light Curves of Type Ia Supernovae Observed with TESS}",
      journal = {\apj},
         year = 2021,
        month = feb,
       volume = {908},
       number = {1},
          eid = {51},
        pages = {51},
          doi = {10.3847/1538-4357/abcd42},
archivePrefix = {arXiv},
       eprint = {1904.02171},
 primaryClass = {astro-ph.SR},
       adsurl = {https://ui.adsabs.harvard.edu/abs/2021ApJ...908...51F}
}

@ARTICLE{shakura73,
       author = {{Shakura}, N.~I. and {Sunyaev}, R.~A.},
        title = "{Black holes in binary systems. Observational appearance.}",
      journal = {\aap},
         year = 1973,
        month = jan,
       volume = {24},
        pages = {337-355},
       adsurl = {https://ui.adsabs.harvard.edu/abs/1973A&A....24..337S}
}

@ARTICLE{morgan10,
       author = {{Morgan}, Christopher W. and {Kochanek}, C.~S. and {Morgan}, Nicholas D. and {Falco}, Emilio E.},
        title = "{The Quasar Accretion Disk Size-Black Hole Mass Relation}",
      journal = {\apj},
         year = 2010,
        month = apr,
       volume = {712},
       number = {2},
        pages = {1129-1136},
          doi = {10.1088/0004-637X/712/2/1129},
archivePrefix = {arXiv},
       eprint = {1002.4160},
 primaryClass = {astro-ph.CO},
       adsurl = {https://ui.adsabs.harvard.edu/abs/2010ApJ...712.1129M}
}

@ARTICLE{kovacevic25,
       author = {{Kovacevic}, Natalie and {Dai}, Xinyu and {Yuk}, Heechan and {J{\"a}rvel{\"a}}, Emilia E. and {Yi}, Tingfeng and {Vallely}, Patrick J. and {Shappee}, Benjamin J. and {Shankar}, Francesco and {Stanek}, K.~Z.},
        title = "{Comparing Optical Variability of Type 1 and Type 2 AGN from the BAT 9 Month Sample Using ASAS-SN and TESS Surveys}",
      journal = {\apj},
         year = 2025,
        month = jun,
       volume = {985},
       number = {2},
          eid = {177},
        pages = {177},
          doi = {10.3847/1538-4357/adcb40},
archivePrefix = {arXiv},
       eprint = {2504.08123},
 primaryClass = {astro-ph.GA},
       adsurl = {https://ui.adsabs.harvard.edu/abs/2025ApJ...985..177K}
}

@ARTICLE{tueller08,
       author = {{Tueller}, J. and {Mushotzky}, R.~F. and {Barthelmy}, S. and {Cannizzo}, J.~K. and {Gehrels}, N. and {Markwardt}, C.~B. and {Skinner}, G.~K. and {Winter}, L.~M.},
        title = "{Swift BAT Survey of AGNs}",
      journal = {\apj},
         year = 2008,
        month = jul,
       volume = {681},
       number = {1},
        pages = {113-127},
          doi = {10.1086/588458},
archivePrefix = {arXiv},
       eprint = {0711.4130},
 primaryClass = {astro-ph},
       adsurl = {https://ui.adsabs.harvard.edu/abs/2008ApJ...681..113T}
}

@ARTICLE{koss17,
       author = {{Koss}, Michael and {Trakhtenbrot}, Benny and {Ricci}, Claudio and {Lamperti}, Isabella and {Oh}, Kyuseok and {Berney}, Simon and {Schawinski}, Kevin and {Balokovi{\'c}}, Mislav and {Baronchelli}, Linda and {Crenshaw}, D. Michael and {Fischer}, Travis and {Gehrels}, Neil and {Harrison}, Fiona and {Hashimoto}, Yasuhiro and {Hogg}, Drew and {Ichikawa}, Kohei and {Masetti}, Nicola and {Mushotzky}, Richard and {Sartori}, Lia and {Stern}, Daniel and {Treister}, Ezequiel and {Ueda}, Yoshihiro and {Veilleux}, Sylvain and {Winter}, Lisa},
        title = "{BAT AGN Spectroscopic Survey. I. Spectral Measurements, Derived Quantities, and AGN Demographics}",
      journal = {\apj},
         year = 2017,
        month = nov,
       volume = {850},
       number = {1},
          eid = {74},
        pages = {74},
          doi = {10.3847/1538-4357/aa8ec9},
archivePrefix = {arXiv},
       eprint = {1707.08123},
 primaryClass = {astro-ph.HE},
       adsurl = {https://ui.adsabs.harvard.edu/abs/2017ApJ...850...74K}
}

@ARTICLE{bennert21,
       author = {{Bennert}, Vardha N. and {Treu}, Tommaso and {Ding}, Xuheng and {Stomberg}, Isak and {Birrer}, Simon and {Snyder}, Tomas and {Malkan}, Matthew A. and {Stephens}, Andrew W. and {Auger}, Matthew W.},
        title = "{A Local Baseline of the Black Hole Mass Scaling Relations for Active Galaxies. IV. Correlations Between M$_{BH}$ and Host Galaxy {\ensuremath{\sigma}}, Stellar Mass, and Luminosity}",
      journal = {\apj},
         year = 2021,
        month = nov,
       volume = {921},
       number = {1},
          eid = {36},
        pages = {36},
          doi = {10.3847/1538-4357/ac151a},
archivePrefix = {arXiv},
       eprint = {2101.10355},
 primaryClass = {astro-ph.GA},
       adsurl = {https://ui.adsabs.harvard.edu/abs/2021ApJ...921...36B}
}

@ARTICLE{zu11,
       author = {{Zu}, Ying and {Kochanek}, C.~S. and {Peterson}, Bradley M.},
        title = "{An Alternative Approach to Measuring Reverberation Lags in Active Galactic Nuclei}",
      journal = {\apj},
         year = 2011,
        month = jul,
       volume = {735},
       number = {2},
          eid = {80},
        pages = {80},
          doi = {10.1088/0004-637X/735/2/80},
archivePrefix = {arXiv},
       eprint = {1008.0641},
 primaryClass = {astro-ph.CO},
       adsurl = {https://ui.adsabs.harvard.edu/abs/2011ApJ...735...80Z}
}

@ARTICLE{bentz15,
       author = {{Bentz}, Misty C. and {Katz}, Sarah},
        title = "{The AGN Black Hole Mass Database}",
      journal = {\pasp},
         year = 2015,
        month = jan,
       volume = {127},
       number = {947},
        pages = {67},
          doi = {10.1086/679601},
archivePrefix = {arXiv},
       eprint = {1411.2596},
 primaryClass = {astro-ph.GA},
       adsurl = {https://ui.adsabs.harvard.edu/abs/2015PASP..127...67B}
}

@ARTICLE{peng06,
       author = {{Peng}, Z. and {Gu}, Q. and {Melnick}, J. and {Zhao}, Y.},
        title = "{The K-band properties of Seyfert 2 galaxies}",
      journal = {\aap},
         year = 2006,
        month = jul,
       volume = {453},
       number = {3},
        pages = {863-868},
          doi = {10.1051/0004-6361:20054664},
archivePrefix = {arXiv},
       eprint = {astro-ph/0603849},
 primaryClass = {astro-ph},
       adsurl = {https://ui.adsabs.harvard.edu/abs/2006A&A...453..863P}
}

@ARTICLE{zhou10,
       author = {{Zhou}, Xin-Lin and {Zhang}, Shuang-Nan and {Wang}, Ding-Xiong and {Zhu}, Ling},
        title = "{Calibrating the Correlation Between Black Hole Mass and X-ray Variability Amplitude: X-ray Only Black Hole Mass Estimates for Active Galactic Nuclei and Ultra-luminous X-ray Sources}",
      journal = {\apj},
         year = 2010,
        month = feb,
       volume = {710},
       number = {1},
        pages = {16-23},
          doi = {10.1088/0004-637X/710/1/16},
archivePrefix = {arXiv},
       eprint = {0912.2636},
 primaryClass = {astro-ph.HE},
       adsurl = {https://ui.adsabs.harvard.edu/abs/2010ApJ...710...16Z}
}

@ARTICLE{bentz09,
       author = {{Bentz}, Misty C. and {Walsh}, Jonelle L. and {Barth}, Aaron J. and {Baliber}, Nairn and {Bennert}, Vardha Nicola and {Canalizo}, Gabriela and {Filippenko}, Alexei V. and {Ganeshalingam}, Mohan and {Gates}, Elinor L. and {Greene}, Jenny E. and {Hidas}, Marton G. and {Hiner}, Kyle D. and {Lee}, Nicholas and {Li}, Weidong and {Malkan}, Matthew A. and {Minezaki}, Takeo and {Sakata}, Yu and {Serduke}, Frank J.~D. and {Silverman}, Jeffrey M. and {Steele}, Thea N. and {Stern}, Daniel and {Street}, Rachel A. and {Thornton}, Carol E. and {Treu}, Tommaso and {Wang}, Xiaofeng and {Woo}, Jong-Hak and {Yoshii}, Yuzuru},
        title = "{The Lick AGN Monitoring Project: Broad-line Region Radii and Black Hole Masses from Reverberation Mapping of H{\ensuremath{\beta}}}",
      journal = {\apj},
         year = 2009,
        month = nov,
       volume = {705},
       number = {1},
        pages = {199-217},
          doi = {10.1088/0004-637X/705/1/199},
archivePrefix = {arXiv},
       eprint = {0908.0003},
 primaryClass = {astro-ph.CO},
       adsurl = {https://ui.adsabs.harvard.edu/abs/2009ApJ...705..199B}
}

@ARTICLE{peterson05,
       author = {{Peterson}, Bradley M. and {Bentz}, Misty C. and {Desroches}, Louis-Benoit and {Filippenko}, Alexei V. and {Ho}, Luis C. and {Kaspi}, Shai and {Laor}, Ari and {Maoz}, Dan and {Moran}, Edward C. and {Pogge}, Richard W. and {Quillen}, Alice C.},
        title = "{Multiwavelength Monitoring of the Dwarf Seyfert 1 Galaxy NGC 4395. I. A Reverberation-based Measurement of the Black Hole Mass}",
      journal = {\apj},
         year = 2005,
        month = oct,
       volume = {632},
       number = {2},
        pages = {799-808},
          doi = {10.1086/444494},
archivePrefix = {arXiv},
       eprint = {astro-ph/0506665},
 primaryClass = {astro-ph},
       adsurl = {https://ui.adsabs.harvard.edu/abs/2005ApJ...632..799P}
}

@ARTICLE{wang07,
       author = {{Wang}, Jian-Min and {Zhang}, En-Peng},
        title = "{The Unified Model of Active Galactic Nuclei. II. Evolutionary Connection}",
      journal = {\apj},
         year = 2007,
        month = may,
       volume = {660},
       number = {2},
        pages = {1072-1092},
          doi = {10.1086/513685},
archivePrefix = {arXiv},
       eprint = {astro-ph/0702279},
 primaryClass = {astro-ph},
       adsurl = {https://ui.adsabs.harvard.edu/abs/2007ApJ...660.1072W}
}

@INPROCEEDINGS{shappee14,
       author = {{Shappee}, Benjamin and {Prieto}, J. and {Stanek}, K.~Z. and {Kochanek}, C.~S. and {Holoien}, T. and {Jencson}, J. and {Basu}, U. and {Beacom}, J.~F. and {Szczygiel}, D. and {Pojmanski}, G. and {Brimacombe}, J. and {Dubberley}, M. and {Elphick}, M. and {Foale}, S. and {Hawkins}, E. and {Mullins}, D. and {Rosing}, W. and {Ross}, R. and {Walker}, Z.},
        title = "{All Sky Automated Survey for SuperNovae (ASAS-SN or ``Assassin'')}",
    booktitle = {American Astronomical Society Meeting Abstracts \#223},
         year = 2014,
       series = {American Astronomical Society Meeting Abstracts},
       volume = {223},
        month = jan,
          eid = {236.03},
        pages = {236.03},
       adsurl = {https://ui.adsabs.harvard.edu/abs/2014AAS...22323603S}
}

@ARTICLE{kochanek17,
       author = {{Kochanek}, C.~S. and {Shappee}, B.~J. and {Stanek}, K.~Z. and {Holoien}, T.~W. -S. and {Thompson}, Todd A. and {Prieto}, J.~L. and {Dong}, Subo and {Shields}, J.~V. and {Will}, D. and {Britt}, C. and {Perzanowski}, D. and {Pojma{\'n}ski}, G.},
        title = "{The All-Sky Automated Survey for Supernovae (ASAS-SN) Light Curve Server v1.0}",
      journal = {\pasp},
         year = 2017,
        month = oct,
       volume = {129},
       number = {980},
        pages = {104502},
          doi = {10.1088/1538-3873/aa80d9},
archivePrefix = {arXiv},
       eprint = {1706.07060},
 primaryClass = {astro-ph.SR},
       adsurl = {https://ui.adsabs.harvard.edu/abs/2017PASP..129j4502K}
}

@ARTICLE{uttley02,
       author = {{Uttley}, P. and {McHardy}, I.~M. and {Papadakis}, I.~E.},
        title = "{Measuring the broad-band power spectra of active galactic nuclei with RXTE}",
      journal = {\mnras},
         year = 2002,
        month = may,
       volume = {332},
       number = {1},
        pages = {231-250},
          doi = {10.1046/j.1365-8711.2002.05298.x},
archivePrefix = {arXiv},
       eprint = {astro-ph/0201134},
 primaryClass = {astro-ph},
       adsurl = {https://ui.adsabs.harvard.edu/abs/2002MNRAS.332..231U}
}

@ARTICLE{macleod10,
       author = {{MacLeod}, C.~L. and {Ivezi{\'c}}, {\v{Z}}. and {Kochanek}, C.~S. and {Koz{\l}owski}, S. and {Kelly}, B. and {Bullock}, E. and {Kimball}, A. and {Sesar}, B. and {Westman}, D. and {Brooks}, K. and {Gibson}, R. and {Becker}, A.~C. and {de Vries}, W.~H.},
        title = "{Modeling the Time Variability of SDSS Stripe 82 Quasars as a Damped Random Walk}",
      journal = {\apj},
         year = 2010,
        month = oct,
       volume = {721},
       number = {2},
        pages = {1014-1033},
          doi = {10.1088/0004-637X/721/2/1014},
archivePrefix = {arXiv},
       eprint = {1004.0276},
 primaryClass = {astro-ph.CO},
       adsurl = {https://ui.adsabs.harvard.edu/abs/2010ApJ...721.1014M}
}

@ARTICLE{morgan08,
       author = {{Morgan}, Christopher W. and {Kochanek}, Christopher. S. and {Dai}, Xinyu and {Morgan}, Nicholas D. and {Falco}, Emilio E.},
        title = "{X-Ray and Optical Microlensing in the Lensed Quasar PG 1115+080}",
      journal = {\apj},
         year = 2008,
        month = dec,
       volume = {689},
       number = {2},
        pages = {755-761},
          doi = {10.1086/592767},
archivePrefix = {arXiv},
       eprint = {0802.1210},
 primaryClass = {astro-ph},
       adsurl = {https://ui.adsabs.harvard.edu/abs/2008ApJ...689..755M}
}

@ARTICLE{dai10,
       author = {{Dai}, X. and {Kochanek}, C.~S. and {Chartas}, G. and {Koz{\l}owski}, S. and {Morgan}, C.~W. and {Garmire}, G. and {Agol}, E.},
        title = "{The Sizes of the X-ray and Optical Emission Regions of RXJ 1131-1231}",
      journal = {\apj},
         year = 2010,
        month = jan,
       volume = {709},
       number = {1},
        pages = {278-285},
          doi = {10.1088/0004-637X/709/1/278},
archivePrefix = {arXiv},
       eprint = {0906.4342},
 primaryClass = {astro-ph.HE},
       adsurl = {https://ui.adsabs.harvard.edu/abs/2010ApJ...709..278D}
}

@ARTICLE{papoutsis24,
       author = {{Papoutsis}, M. and {Papadakis}, I.~E. and {Panagiotou}, C. and {Dov{\v{c}}iak}, M. and {Kammoun}, E.},
        title = "{X-ray reverberation as an explanation for UV/optical variability in nearby Seyferts}",
      journal = {\aap},
         year = 2024,
        month = nov,
       volume = {691},
          eid = {A60},
        pages = {A60},
          doi = {10.1051/0004-6361/202348603},
archivePrefix = {arXiv},
       eprint = {2409.10417},
 primaryClass = {astro-ph.HE},
       adsurl = {https://ui.adsabs.harvard.edu/abs/2024A&A...691A..60P}
}

@ARTICLE{shappee14b,
       author = {{Shappee}, B.~J. and {Prieto}, J.~L. and {Grupe}, D. and {Kochanek}, C.~S. and {Stanek}, K.~Z. and {De Rosa}, G. and {Mathur}, S. and {Zu}, Y. and {Peterson}, B.~M. and {Pogge}, R.~W. and {Komossa}, S. and {Im}, M. and {Jencson}, J. and {Holoien}, T.~W. -S. and {Basu}, U. and {Beacom}, J.~F. and {Szczygie{\l}}, D.~M. and {Brimacombe}, J. and {Adams}, S. and {Campillay}, A. and {Choi}, C. and {Contreras}, C. and {Dietrich}, M. and {Dubberley}, M. and {Elphick}, M. and {Foale}, S. and {Giustini}, M. and {Gonzalez}, C. and {Hawkins}, E. and {Howell}, D.~A. and {Hsiao}, E.~Y. and {Koss}, M. and {Leighly}, K.~M. and {Morrell}, N. and {Mudd}, D. and {Mullins}, D. and {Nugent}, J.~M. and {Parrent}, J. and {Phillips}, M.~M. and {Pojmanski}, G. and {Rosing}, W. and {Ross}, R. and {Sand}, D. and {Terndrup}, D.~M. and {Valenti}, S. and {Walker}, Z. and {Yoon}, Y.},
        title = "{The Man behind the Curtain: X-Rays Drive the UV through NIR Variability in the 2013 Active Galactic Nucleus Outburst in NGC 2617}",
      journal = {\apj},
         year = 2014,
        month = jun,
       volume = {788},
       number = {1},
          eid = {48},
        pages = {48},
          doi = {10.1088/0004-637X/788/1/48},
archivePrefix = {arXiv},
       eprint = {1310.2241},
 primaryClass = {astro-ph.HE},
       adsurl = {https://ui.adsabs.harvard.edu/abs/2014ApJ...788...48S}
}

@ARTICLE{pounds90,
       author = {{Pounds}, K.~A. and {Nandra}, K. and {Stewart}, G.~C. and {George}, I.~M. and {Fabian}, A.~C.},
        title = "{X-ray reflection from cold matter in the nuclei of active galaxies}",
      journal = {\nat},
         year = 1990,
        month = mar,
       volume = {344},
       number = {6262},
        pages = {132-133},
          doi = {10.1038/344132a0},
       adsurl = {https://ui.adsabs.harvard.edu/abs/1990Natur.344..132P}
}

@ARTICLE{george91,
       author = {{George}, I.~M. and {Fabian}, A.~C.},
        title = "{X-ray reflection from cold matter in Active Galactic Nuclei and X-ray binaries.}",
      journal = {\mnras},
         year = 1991,
        month = mar,
       volume = {249},
        pages = {352},
          doi = {10.1093/mnras/249.2.352},
       adsurl = {https://ui.adsabs.harvard.edu/abs/1991MNRAS.249..352G}
}

@ARTICLE{kawaguchi98,
       author = {{Kawaguchi}, T. and {Mineshige}, S. and {Umemura}, M. and {Turner}, Edwin L.},
        title = "{Optical Variability in Active Galactic Nuclei: Starbursts or Disk Instabilities?}",
      journal = {\apj},
         year = 1998,
        month = sep,
       volume = {504},
       number = {2},
        pages = {671-679},
          doi = {10.1086/306105},
archivePrefix = {arXiv},
       eprint = {astro-ph/9712006},
 primaryClass = {astro-ph},
       adsurl = {https://ui.adsabs.harvard.edu/abs/1998ApJ...504..671K}
}

@ARTICLE{trevese02,
       author = {{Tr{\`e}vese}, Dario and {Vagnetti}, Fausto},
        title = "{Quasar Spectral Slope Variability in the Optical Band}",
      journal = {\apj},
         year = 2002,
        month = jan,
       volume = {564},
       number = {2},
        pages = {624-630},
          doi = {10.1086/324541},
archivePrefix = {arXiv},
       eprint = {astro-ph/0110075},
 primaryClass = {astro-ph},
       adsurl = {https://ui.adsabs.harvard.edu/abs/2002ApJ...564..624T}
}

@ARTICLE{kelly09,
       author = {{Kelly}, Brandon C. and {Bechtold}, Jill and {Siemiginowska}, Aneta},
        title = "{Are the Variations in Quasar Optical Flux Driven by Thermal Fluctuations?}",
      journal = {\apj},
         year = 2009,
        month = jun,
       volume = {698},
       number = {1},
        pages = {895-910},
          doi = {10.1088/0004-637X/698/1/895},
archivePrefix = {arXiv},
       eprint = {0903.5315},
 primaryClass = {astro-ph.CO},
       adsurl = {https://ui.adsabs.harvard.edu/abs/2009ApJ...698..895K}
}

@ARTICLE{gonzalezmartin12,
       author = {{Gonz{\'a}lez-Mart{\'\i}n}, O. and {Vaughan}, S.},
        title = "{X-ray variability of 104 active galactic nuclei. XMM-Newton power-spectrum density profiles}",
      journal = {\aap},
         year = 2012,
        month = aug,
       volume = {544},
          eid = {A80},
        pages = {A80},
          doi = {10.1051/0004-6361/201219008},
archivePrefix = {arXiv},
       eprint = {1205.4255},
 primaryClass = {astro-ph.HE},
       adsurl = {https://ui.adsabs.harvard.edu/abs/2012A&A...544A..80G}
}

@ARTICLE{lyndenbell69,
       author = {{Lynden-Bell}, D.},
        title = "{Galactic Nuclei as Collapsed Old Quasars}",
      journal = {\nat},
         year = 1969,
        month = aug,
       volume = {223},
       number = {5207},
        pages = {690-694},
          doi = {10.1038/223690a0},
       adsurl = {https://ui.adsabs.harvard.edu/abs/1969Natur.223..690L}
}

@ARTICLE{peterson93,
       author = {{Peterson}, Bradley M.},
        title = "{Reverberation Mapping of Active Galactic Nuclei}",
      journal = {\pasp},
         year = 1993,
        month = mar,
       volume = {105},
        pages = {247},
          doi = {10.1086/133140},
       adsurl = {https://ui.adsabs.harvard.edu/abs/1993PASP..105..247P}
}

@ARTICLE{yang24,
       author = {{Yang}, Yujian and {Ma}, Bo and {Chen}, Chen},
        title = "{Studying Intra-Night Optical Variability of AGNs Using the TESS Survey Data}",
      journal = {Universe},
         year = 2024,
        month = nov,
       volume = {10},
       number = {12},
          eid = {434},
        pages = {434},
          doi = {10.3390/universe10120434},
       adsurl = {https://ui.adsabs.harvard.edu/abs/2024Univ...10..434Y}
}

@ARTICLE{dingler24,
       author = {{Dingler}, Ryne and {Smith}, Krista Lynne},
        title = "{Optical Variability Properties of Southern TESS Blazars}",
      journal = {\apj},
         year = 2024,
        month = sep,
       volume = {973},
       number = {1},
          eid = {10},
        pages = {10},
          doi = {10.3847/1538-4357/ad4f87},
archivePrefix = {arXiv},
       eprint = {2406.10346},
 primaryClass = {astro-ph.HE},
       adsurl = {https://ui.adsabs.harvard.edu/abs/2024ApJ...973...10D}
}

@ARTICLE{paolillo25,
       author = {{Paolillo}, Maurizio and {Papadakis}, Iossif},
        title = "{Continuum optical-UV and X-ray variability of AGN: current results and future challenges}",
      journal = {Nuovo Cimento Rivista Serie},
         year = 2025,
        month = aug,
          doi = {10.1007/s40766-025-00072-5},
archivePrefix = {arXiv},
       eprint = {2506.23899},
 primaryClass = {astro-ph.HE},
       adsurl = {https://ui.adsabs.harvard.edu/abs/2025NCimR.tmp...19P}
}

@ARTICLE{edelson99,
       author = {{Edelson}, Rick and {Nandra}, Kirpal},
        title = "{A Cutoff in the X-Ray Fluctuation Power Density Spectrum of the Seyfert 1 Galaxy NGC 3516}",
      journal = {\apj},
         year = 1999,
        month = apr,
       volume = {514},
       number = {2},
        pages = {682-690},
          doi = {10.1086/306980},
archivePrefix = {arXiv},
       eprint = {astro-ph/9810481},
 primaryClass = {astro-ph},
       adsurl = {https://ui.adsabs.harvard.edu/abs/1999ApJ...514..682E}
}

@ARTICLE{smith18,
       author = {{Smith}, Krista Lynne and {Mushotzky}, Richard F. and {Boyd}, Patricia T. and {Malkan}, Matt and {Howell}, Steve B. and {Gelino}, Dawn M.},
        title = "{The Kepler Light Curves of AGN: A Detailed Analysis}",
      journal = {\apj},
         year = 2018,
        month = apr,
       volume = {857},
       number = {2},
          eid = {141},
        pages = {141},
          doi = {10.3847/1538-4357/aab88d},
archivePrefix = {arXiv},
       eprint = {1803.06436},
 primaryClass = {astro-ph.HE},
       adsurl = {https://ui.adsabs.harvard.edu/abs/2018ApJ...857..141S}
}

@INPROCEEDINGS{czerny06,
       author = {{Czerny}, B.},
        title = "{The Role of the Accretion Disk in AGN Variability}",
    booktitle = {AGN Variability from X-Rays to Radio Waves},
         year = 2006,
       editor = {{Gaskell}, C. Martin and {McHardy}, Ian M. and {Peterson}, Bradley M. and {Sergeev}, Sergey G.},
       series = {Astronomical Society of the Pacific Conference Series},
       volume = {360},
        month = dec,
        pages = {265},
       adsurl = {https://ui.adsabs.harvard.edu/abs/2006ASPC..360..265C}
}

@ARTICLE{goyal12,
       author = {{Goyal}, A. and {Gopal-Krishna} and {Wiita}, P.~J. and {Anupama}, G.~C. and {Sahu}, D.~K. and {Sagar}, R. and {Joshi}, S.},
        title = "{Intra-night optical variability of core dominated radio quasars: the role of optical polarization}",
      journal = {\aap},
         year = 2012,
        month = aug,
       volume = {544},
          eid = {A37},
        pages = {A37},
          doi = {10.1051/0004-6361/201218888},
archivePrefix = {arXiv},
       eprint = {1205.2324},
 primaryClass = {astro-ph.HE},
       adsurl = {https://ui.adsabs.harvard.edu/abs/2012A&A...544A..37G}
}

@ARTICLE{goyal13,
       author = {{Goyal}, Arti and {Gopal-Krishna}, Paul J., Wiita and {Stalin}, C.~S. and {Sagar}, Ram},
        title = "{Improved characterization of intranight optical variability of prominent AGN classes}",
      journal = {\mnras},
         year = 2013,
        month = oct,
       volume = {435},
       number = {2},
        pages = {1300-1312},
          doi = {10.1093/mnras/stt1373},
archivePrefix = {arXiv},
       eprint = {1307.5831},
 primaryClass = {astro-ph.HE},
       adsurl = {https://ui.adsabs.harvard.edu/abs/2013MNRAS.435.1300G}
}

@ARTICLE{kumar15,
       author = {{Kumar}, Parveen and {Gopal-Krishna}, Hum, Chand},
        title = "{Intranight optical variability of radio-quiet weak emission line quasars - III}",
      journal = {\mnras},
         year = 2015,
        month = apr,
       volume = {448},
       number = {2},
        pages = {1463-1470},
          doi = {10.1093/mnras/stv060},
archivePrefix = {arXiv},
       eprint = {1501.02100},
 primaryClass = {astro-ph.HE},
       adsurl = {https://ui.adsabs.harvard.edu/abs/2015MNRAS.448.1463K}
}

@ARTICLE{aranzana18,
       author = {{Aranzana}, E. and {K{\"o}rding}, E. and {Uttley}, P. and {Scaringi}, S. and {Bloemen}, S.},
        title = "{Short time-scale optical variability properties of the largest AGN sample observed with Kepler/K2}",
      journal = {\mnras},
         year = 2018,
        month = may,
       volume = {476},
       number = {2},
        pages = {2501-2515},
          doi = {10.1093/mnras/sty413},
archivePrefix = {arXiv},
       eprint = {1802.08058},
 primaryClass = {astro-ph.HE},
       adsurl = {https://ui.adsabs.harvard.edu/abs/2018MNRAS.476.2501A}
}

@ARTICLE{skrutskie06,
       author = {{Skrutskie}, M.~F. and {Cutri}, R.~M. and {Stiening}, R. and {Weinberg}, M.~D. and {Schneider}, S. and {Carpenter}, J.~M. and {Beichman}, C. and {Capps}, R. and {Chester}, T. and {Elias}, J. and {Huchra}, J. and {Liebert}, J. and {Lonsdale}, C. and {Monet}, D.~G. and {Price}, S. and {Seitzer}, P. and {Jarrett}, T. and {Kirkpatrick}, J.~D. and {Gizis}, J.~E. and {Howard}, E. and {Evans}, T. and {Fowler}, J. and {Fullmer}, L. and {Hurt}, R. and {Light}, R. and {Kopan}, E.~L. and {Marsh}, K.~A. and {McCallon}, H.~L. and {Tam}, R. and {Van Dyk}, S. and {Wheelock}, S.},
        title = "{The Two Micron All Sky Survey (2MASS)}",
      journal = {\aj},
         year = 2006,
        month = feb,
       volume = {131},
       number = {2},
        pages = {1163-1183},
          doi = {10.1086/498708},
       adsurl = {https://ui.adsabs.harvard.edu/abs/2006AJ....131.1163S}
}

@ARTICLE{jarrett00,
       author = {{Jarrett}, T.~H. and {Chester}, T. and {Cutri}, R. and {Schneider}, S. and {Skrutskie}, M. and {Huchra}, J.~P.},
        title = "{2MASS Extended Source Catalog: Overview and Algorithms}",
      journal = {\aj},
         year = 2000,
        month = may,
       volume = {119},
       number = {5},
        pages = {2498-2531},
          doi = {10.1086/301330},
archivePrefix = {arXiv},
       eprint = {astro-ph/0004318},
 primaryClass = {astro-ph},
       adsurl = {https://ui.adsabs.harvard.edu/abs/2000AJ....119.2498J}
}

@ARTICLE{burke20,
       author = {{Burke}, Colin J. and {Shen}, Yue and {Chen}, Yu-Ching and {Scaringi}, Simone and {Faucher-Giguere}, Claude-Andre and {Liu}, Xin and {Yang}, Qian},
        title = "{Optical Variability of the Dwarf AGN NGC 4395 from the Transiting Exoplanet Survey Satellite}",
      journal = {\apj},
         year = 2020,
        month = aug,
       volume = {899},
       number = {2},
          eid = {136},
        pages = {136},
          doi = {10.3847/1538-4357/aba3ce},
archivePrefix = {arXiv},
       eprint = {2005.04491},
 primaryClass = {astro-ph.GA},
       adsurl = {https://ui.adsabs.harvard.edu/abs/2020ApJ...899..136B}
}

@ARTICLE{papadakis04,
       author = {{Papadakis}, I.~E.},
        title = "{The scaling of the X-ray variability with black hole mass in active galactic nuclei}",
      journal = {\mnras},
         year = 2004,
        month = feb,
       volume = {348},
       number = {1},
        pages = {207-213},
          doi = {10.1111/j.1365-2966.2004.07351.x},
archivePrefix = {arXiv},
       eprint = {astro-ph/0311016},
 primaryClass = {astro-ph},
       adsurl = {https://ui.adsabs.harvard.edu/abs/2004MNRAS.348..207P}
}

@ARTICLE{lu01,
       author = {{Lu}, Youjun and {Yu}, Qingjuan},
        title = "{The relationship between X-ray variability and the central black hole mass}",
      journal = {\mnras},
         year = 2001,
        month = jun,
       volume = {324},
       number = {3},
        pages = {653-658},
          doi = {10.1046/j.1365-8711.2001.04344.x},
archivePrefix = {arXiv},
       eprint = {astro-ph/0106292},
 primaryClass = {astro-ph},
       adsurl = {https://ui.adsabs.harvard.edu/abs/2001MNRAS.324..653L}
}

@ARTICLE{wang22,
       author = {{Wang}, Hong-Tao and {Su}, Yan-Ping and {Ge}, Xue and {Chen}, Yong-Yun and {Yu}, Xiao-Ling},
        title = "{The Variability of the Narrow-line Seyfert 1 Galaxies from the Pan-STARRS's View}",
      journal = {Research in Astronomy and Astrophysics},
         year = 2022,
        month = jan,
       volume = {22},
       number = {1},
          eid = {015014},
        pages = {015014},
          doi = {10.1088/1674-4527/ac3895},
archivePrefix = {arXiv},
       eprint = {2201.08007},
 primaryClass = {astro-ph.GA},
       adsurl = {https://ui.adsabs.harvard.edu/abs/2022RAA....22a5014W}
}

@ARTICLE{romero99,
       author = {{Romero}, G.~E. and {Cellone}, S.~A. and {Combi}, J.~A.},
        title = "{Optical microvariability of southern AGNs}",
      journal = {\aaps},
         year = 1999,
        month = mar,
       volume = {135},
        pages = {477-486},
          doi = {10.1051/aas:1999184},
       adsurl = {https://ui.adsabs.harvard.edu/abs/1999A&AS..135..477R}
}

@ARTICLE{markowitz03,
       author = {{Markowitz}, A. and {Edelson}, R. and {Vaughan}, S. and {Uttley}, P. and {George}, I.~M. and {Griffiths}, R.~E. and {Kaspi}, S. and {Lawrence}, A. and {McHardy}, I. and {Nandra}, K. and {Pounds}, K. and {Reeves}, J. and {Schurch}, N. and {Warwick}, R.},
        title = "{X-Ray Fluctuation Power Spectral Densities of Seyfert 1 Galaxies}",
      journal = {\apj},
         year = 2003,
        month = aug,
       volume = {593},
       number = {1},
        pages = {96-114},
          doi = {10.1086/375330},
archivePrefix = {arXiv},
       eprint = {astro-ph/0303273},
 primaryClass = {astro-ph},
       adsurl = {https://ui.adsabs.harvard.edu/abs/2003ApJ...593...96M}
}

@ARTICLE{tarrant25,
       author = {{Tarrant}, Ashley and {Hinkle}, Jason and {Shappee}, Benjamin and {Kochanek}, Christopher and {Hey}, Daniel and {Auge}, Connor and {Payne}, Anna and {Bolish}, Michael and {Yuk}, Heechan and {Dai}, Xinyu and {Auchettl}, Katie and {Thompson}, Todd and {Treiber}, Helena},
        title = "{The AGN Optical Variability Fundamental Plane}",
      journal = {arXiv e-prints},
         year = 2025,
        month = jan,
          eid = {arXiv:2501.12444},
        pages = {arXiv:2501.12444},
          doi = {10.48550/arXiv.2501.12444},
archivePrefix = {arXiv},
       eprint = {2501.12444},
 primaryClass = {astro-ph.GA},
       adsurl = {https://ui.adsabs.harvard.edu/abs/2025arXiv250112444T}
}

@ARTICLE{fabian15,
       author = {{Fabian}, A.~C. and {Lohfink}, A. and {Kara}, E. and {Parker}, M.~L. and {Vasudevan}, R. and {Reynolds}, C.~S.},
        title = "{Properties of AGN coronae in the NuSTAR era}",
      journal = {\mnras},
         year = 2015,
        month = aug,
       volume = {451},
       number = {4},
        pages = {4375-4383},
          doi = {10.1093/mnras/stv1218},
archivePrefix = {arXiv},
       eprint = {1505.07603},
 primaryClass = {astro-ph.HE},
       adsurl = {https://ui.adsabs.harvard.edu/abs/2015MNRAS.451.4375F}
}

@ARTICLE{pozonunez15,
       author = {{Pozo Nu{\~n}ez}, F. and {Ramolla}, M. and {Westhues}, C. and {Haas}, M. and {Chini}, R. and {Steenbrugge}, K. and {Barr Dom{\'\i}nguez}, A. and {Kaderhandt}, L. and {Hackstein}, M. and {Kollatschny}, W. and {Zetzl}, M. and {Hodapp}, K.~W. and {Murphy}, M.},
        title = "{The broad-line region and dust torus size of the Seyfert 1 galaxy PGC 50427}",
      journal = {\aap},
         year = 2015,
        month = apr,
       volume = {576},
          eid = {A73},
        pages = {A73},
          doi = {10.1051/0004-6361/201525910},
archivePrefix = {arXiv},
       eprint = {1502.06771},
 primaryClass = {astro-ph.GA},
       adsurl = {https://ui.adsabs.harvard.edu/abs/2015A&A...576A..73P}
}

@ARTICLE{dogruel20,
       author = {{Dogruel}, Mustafa Burak and {Dai}, Xinyu and {Guerras}, Eduardo and {Cornachione}, Matthew and {Morgan}, Christopher W.},
        title = "{X-Ray Monitoring of Gravitationally Lensed Radio-loud Quasars with Chandra}",
      journal = {\apj},
         year = 2020,
        month = may,
       volume = {894},
       number = {2},
          eid = {153},
        pages = {153},
          doi = {10.3847/1538-4357/ab879b},
archivePrefix = {arXiv},
       eprint = {1909.06741},
 primaryClass = {astro-ph.CO},
       adsurl = {https://ui.adsabs.harvard.edu/abs/2020ApJ...894..153D}
}

@ARTICLE{rani25,
       author = {{Rani}, B. and {Kim}, Jungeun and {Papadakis}, I. and {Gendreau}, K.~C. and {Masterson}, M. and {Hamaguchi}, K. and {Kara}, E. and {Lee}, S. -S. and {Mushotzky}, R.},
        title = "{High-frequency Power Spectrum of Active Galactic Nucleus NGC 4051 Revealed by NICER}",
      journal = {\apjl},
         year = 2025,
        month = mar,
       volume = {981},
       number = {1},
          eid = {L18},
        pages = {L18},
          doi = {10.3847/2041-8213/adace8},
       adsurl = {https://ui.adsabs.harvard.edu/abs/2025ApJ...981L..18R}
}

@ARTICLE{jha22,
       author = {{Jha}, Vivek Kumar and {Joshi}, Ravi and {Chand}, Hum and {Wu}, Xue-Bing and {Ho}, Luis C. and {Rastogi}, Shantanu and {Ma}, Qinchun},
        title = "{Accretion disc sizes from continuum reverberation mapping of AGN selected from the ZTF survey}",
      journal = {\mnras},
         year = 2022,
        month = apr,
       volume = {511},
       number = {2},
        pages = {3005-3016},
          doi = {10.1093/mnras/stac109},
archivePrefix = {arXiv},
       eprint = {2109.05036},
 primaryClass = {astro-ph.GA},
       adsurl = {https://ui.adsabs.harvard.edu/abs/2022MNRAS.511.3005J}
}

@ARTICLE{dexter12,
       author = {{Dexter}, Jason and {Quataert}, Eliot},
        title = "{Inhomogeneous accretion discs and the soft states of black hole X-ray binaries}",
      journal = {\mnras},
         year = 2012,
        month = oct,
       volume = {426},
       number = {1},
        pages = {L71-L75},
          doi = {10.1111/j.1745-3933.2012.01328.x},
archivePrefix = {arXiv},
       eprint = {1206.0739},
 primaryClass = {astro-ph.HE},
       adsurl = {https://ui.adsabs.harvard.edu/abs/2012MNRAS.426L..71D}
}

@ARTICLE{chiaberge11,
       author = {{Chiaberge}, Marco and {Marconi}, Alessandro},
        title = "{On the origin of radio loudness in active galactic nuclei and its relationship with the properties of the central supermassive black hole}",
      journal = {\mnras},
         year = 2011,
        month = sep,
       volume = {416},
       number = {2},
        pages = {917-926},
          doi = {10.1111/j.1365-2966.2011.19079.x},
archivePrefix = {arXiv},
       eprint = {1105.4889},
 primaryClass = {astro-ph.CO},
       adsurl = {https://ui.adsabs.harvard.edu/abs/2011MNRAS.416..917C}
}

@ARTICLE{kozlowski16,
       author = {{Koz{\l}owski}, Szymon},
        title = "{Revisiting Stochastic Variability of AGNs with Structure Functions}",
      journal = {\apj},
         year = 2016,
        month = aug,
       volume = {826},
       number = {2},
          eid = {118},
        pages = {118},
          doi = {10.3847/0004-637X/826/2/118},
archivePrefix = {arXiv},
       eprint = {1604.05858},
 primaryClass = {astro-ph.GA},
       adsurl = {https://ui.adsabs.harvard.edu/abs/2016ApJ...826..118K}
}

@ARTICLE{bauer09,
       author = {{Bauer}, Anne and {Baltay}, Charles and {Coppi}, Paolo and {Ellman}, Nancy and {Jerke}, Jonathan and {Rabinowitz}, David and {Scalzo}, Richard},
        title = "{Quasar Optical Variability in the Palomar-QUEST Survey}",
      journal = {\apj},
         year = 2009,
        month = may,
       volume = {696},
       number = {2},
        pages = {1241-1256},
          doi = {10.1088/0004-637X/696/2/1241},
archivePrefix = {arXiv},
       eprint = {0902.4103},
 primaryClass = {astro-ph.CO},
       adsurl = {https://ui.adsabs.harvard.edu/abs/2009ApJ...696.1241B}
}

@ARTICLE{middei17,
       author = {{Middei}, R. and {Vagnetti}, F. and {Bianchi}, S. and {La Franca}, F. and {Paolillo}, M. and {Ursini}, F.},
        title = "{A long-term study of AGN X-ray variability . Structure function analysis on a ROSAT-XMM quasar sample}",
      journal = {\aap},
         year = 2017,
        month = mar,
       volume = {599},
          eid = {A82},
        pages = {A82},
          doi = {10.1051/0004-6361/201629940},
archivePrefix = {arXiv},
       eprint = {1612.08547},
 primaryClass = {astro-ph.HE},
       adsurl = {https://ui.adsabs.harvard.edu/abs/2017A&A...599A..82M}
}

\begin{appendix}

\section{Properties of the sample}

Here, we present the properties of the sample of 47 Seyfert 1 galaxies used in this study, including redshift, $V$-band luminosity, black hole mass, and TESS sectors used for light curve analysis (Table \ref{tesssecs}).

\begin{table*}[]
\caption{Summary of the radio-quiet Seyfert 1 galaxy sample}
\centering
\begin{tabular}{lcccl}
\hline\hline
Name & Redshift & $\log(L_V/(\textrm{erg s}^{-1}))$ & $\log(M_{\textrm{BH}}/M_{\odot})$ & TESS sectors used \\ \hline
NGC 235A                & 0.0229 & $44.08\pm0.02$ & $8.6 \pm 0.3 ^1 $   & 3 \\
ESO 113-G010            & 0.0257 & $43.94\pm0.03$ & $7.8 \pm 0.5 ^* $   & 1, 2, 28, 29 \\
Fairall 9               & 0.0461 & $44.68\pm0.07$ & $8.3 \pm 0.1 ^2 $   & 2, 28, 29 \\
Mrk 590                 & 0.0264 & $44.20\pm0.02$ & $7.57 \pm 0.07 ^3 $ & 4 \\
NGC 931                 & 0.0166 & $43.81\pm0.04$ & $7.29 \pm 0.3 ^1 $  & 18 \\
ESO 198-G024            & 0.0455 & $44.21\pm0.19$ & $8 \pm 0.5 ^* $     & 2, 3 \\
1H 0419-577             & 0.1041 & $45.15\pm0.13$ & $8.07 \pm 0.3 ^1 $  & 2, 3, 4, 5 \\
1RXS J045205.0+493248   & 0.0277 & $44.11\pm0.13$ & $8.12 \pm 0.3 ^1 $  & 19 \\
ARK 120                 & 0.0327 & $44.51\pm0.10$ & $8.07 \pm 0.06 ^3 $ & 5 \\
ESO 362-G018            & 0.0124 & $43.68\pm0.03$ & $7.42 \pm 0.3 ^1 $  & 5, 6 \\
MCG +08-11-011          & 0.0205 & $44.09\pm0.07$ & $7.62 \pm 0.3 ^1 $  & 19 \\
EXO 055620-3820.2       & 0.0339 & $43.99\pm0.05$ & $6.87 \pm 0.3 ^1 $  & 5 \\
IRAS 05589+2828         & 0.0329 & $44.38\pm0.12$ & $7.79 \pm 0.3 ^1 $  & 43, 44, 45 \\
ESO 490-G026            & 0.0249 & $44.07\pm0.03$ & $7.9 \pm 0.5 ^* $   & 6, 7 \\
Mrk 79                  & 0.0222 & $43.96\pm0.05$ & $7.61 \pm 0.13 ^3 $ & 20 \\
IGR J07597-3842         & 0.0400 & $44.18\pm0.20$ & $8.45 \pm 0.3 ^1 $  & 7, 8 \\
2MASX J09043699+5536025 & 0.0372 & $43.82\pm0.10$ & $7.09 \pm 0.3 ^1 $  & 20, 21 \\
IRAS 09149-6206         & 0.0573 & $45.01\pm0.05$ & $8.58 \pm 0.3 ^1 $  & 9, 10 \\
MCG +04-22-042          & 0.0331 & $44.12\pm0.10$ & $7.34 \pm 0.3 ^1 $  & 21 \\
Mrk 110                 & 0.0353 & $44.18\pm0.14$ & $7.29 \pm 0.1 ^3 $  & 21 \\
ESO 434-G040            & 0.0085 & $43.40\pm0.02$ & $7.57 \pm 0.25 ^4 $ & 9, 35 \\
NGC 3227                & 0.0038 & $42.92\pm0.03$ & $6.8 \pm 0.1 ^2 $   & 45, 46, 48 \\
2MASX J10384520-4946531 & 0.0600 & $44.51\pm0.14$ & $8.36 \pm 0.3 ^1 $  & 9, 10 \\
NGC 3516                & 0.0088 & $43.74\pm0.02$ & $7.4 \pm 0.05 ^2 $  & 14, 20, 21, 41, 47, 48 \\
NGC 3783                & 0.0097 & $43.82\pm0.11$ & $7.37 \pm 0.08 ^2 $ & 10, 36, 37 \\
SBS 1136+594            & 0.0612 & $44.26\pm0.14$ & $7.98 \pm 0.3 ^1 $  & 14, 15 \\
UGC 06728               & 0.0065 & $42.79\pm0.04$ & $5.66 \pm 0.3 ^1 $  & 14 \\
2MASX J11454045-1827149 & 0.0329 & $44.21\pm0.10$ & $7.31 \pm 0.3 ^1 $  & 9 \\
NGC 4051                & 0.0023 & $42.40\pm0.03$ & $6.1 \pm 0.1 ^2 $   & 22, 49 \\
NGC 4151                & 0.0033 & $43.11\pm0.03$ & $7.55 \pm 0.05 ^2 $ & 49 \\
Mrk 766                 & 0.0129 & $43.59\pm0.03$ & $6.2 \pm 0.3 ^5 $   & 22 \\
NGC 4395                & 0.0011 & $40.93\pm0.04$ & $5.4 \pm 0.1 ^6 $   & 22, 49 \\
NGC 4593                & 0.0083 & $43.55\pm0.03$ & $6.88 \pm 0.09 ^3 $ & 46 \\
SBS 1301+540            & 0.0301 & $43.82\pm0.15$ & $7.73 \pm 0.3 ^1 $  & 15, 16 \\
MCG -06-30-015          & 0.0077 & $43.28\pm0.03$ & $6.3 \pm 0.4 ^7 $   & 37 \\
Mrk 279                 & 0.0305 & $44.13\pm0.08$ & $7.43 \pm 0.12 ^3 $ & 14, 15 \\
NGC 5548                & 0.0172 & $44.00\pm0.07$ & $7.72 \pm 0.02 ^2 $ & 23, 50 \\
ESO 511-G030            & 0.0224 & $43.99\pm0.04$ & $7.13 \pm 0.3 ^1 $  & 11 \\
Mrk 841                 & 0.0364 & $44.26\pm0.12$ & $7.81 \pm 0.3 ^1 $  & 51 \\
Mrk 290                 & 0.0302 & $43.94\pm0.07$ & $7.28 \pm 0.06 ^3 $ & 15, 16 \\
2MASX J16481523-3035037 & 0.0313 & $44.05\pm0.09$ & $7.9 \pm 0.5 ^* $   & 12 \\
NGC 6814                & 0.0052 & $43.19\pm0.04$ & $7.04 \pm 0.06 ^3 $ & 54 \\
NGC 6860                & 0.0151 & $43.86\pm0.04$ & $7.6 \pm 0.5 ^8 $   & 13, 27 \\
IGR J21277+5656         & 0.0149 & $45.00\pm0.03$ & $8.9 \pm 0.5 ^* $   & 15, 16, 17 \\
NGC 7213                & 0.0058 & $43.73\pm0.01$ & $7.6 \pm 0.5 ^* $   & 28 \\
MR 2251-178             & 0.0640 & $44.91\pm0.12$ & $8.44 \pm 0.3 ^1 $  & 2 \\
Mrk 926                 & 0.0470 & $44.54\pm0.07$ & $8.55 \pm 0.3 ^1 $  & 42 \\ \hline
\end{tabular}
\tablefoot{The redshift measurements are retrieved from NED. $L_V$ is computed from ASAS-SN light curves. Black hole mass references: $^1$\citet{koss17}; $^2$\citet{zu11}; $^3$\citet{bentz15}; $^4$\citet{peng06}; $^5$\citet{bentz09}; $^6$\citet{peterson05}; $^7$\citet{zhou10}; $^8$\citet{wang07}; $^*$Measured from spheroid luminosity estimate \citep{bennert21,yuk22}.}
\label{tesssecs}
\end{table*}

\end{appendix}

\end{document}